\documentclass[author,genTeX,type1rest]{nrc1}
\usepackage{graphicx}
\usepackage{amsmath}
\usepackage{amssymb}
\usepackage{cite}
\usepackage[french,english]{babel}

\def\la{\mathrel{\hbox{\rlap{\hbox{\lower4pt\hbox{$\sim$}}}\hbox{$<$}}}}
\def\ga{\mathrel{\hbox{\rlap{\hbox{\lower4pt\hbox{$\sim$}}}\hbox{$>$}}}}

\newcommand{\chandra}{{\it Chandra}}
\newcommand{\rosat}{{\it ROSAT\,}}
\newcommand{\asca}{{\it ASCA}}

\newcommand{\xmm}{{\it XMM-Newton}}
\newcommand{\euve}{{\it EUVE}}
\newcommand{\swift}{{\it Swift}}
\newcommand{\suzaku}{{\it Suzaku}}

 \newcommand{\ka}{K$\alpha$}
\newcommand{\kb}{K$\beta$}

\newcommand{\lya}{Ly$\alpha$}

\begin{document}

\title{EBIT Charge-Exchange Measurements and Astrophysical Applications}
\author{B.J.\ Wargelin}
\address[CfA]{Harvard-Smithsonian Center for Astrophysics,
	60 Garden Street, Cambridge, MA 02138, USA}
\correspond{bwargelin@cfa.harvard.edu}
\author{P.\ Beiersdorfer}
\address[LLNL]{Department of Physics, Lawrence Livermore National Laboratory,
	Livermore, CA 94550, USA}
\author{G.V.\ Brown}
\address[LLNL]
\shortauthor{Wargelin, et al.}
\maketitle



\begin{abstract}

The past decade has seen a surge of interest in astrophysical
charge exchange (CX).  The impetus was the discovery of X-ray emission
from comets in 1996, soon followed by the observation of CX
emission in planetary atmospheres and throughout the heliosphere.
Geocoronal and heliospheric CX are now recognized to contribute
a considerable fraction of the soft X-ray background, and
stellar-wind charge exchange is expected to occur in the astrospheres
surrounding many stars.
CX may also contribute to X-ray line emission in
supernova remnants, the Galactic Center, and the Galactic Ridge.
This article summarizes the key aspects of CX X-ray emission and its
astrophysical relevance, and reviews related
laboratory measurements and theoretical predictions 
with particular attention to spectroscopy experiments 
conducted on electron beam ion traps.


\end{abstract}

\section{Introduction}
\label{sec:intro}

In addition to electron impact excitation, an atomic process 
known to every astronomer and physicist,
there are other line excitation processes including
radiative recombination cascades, 
inner-shell ionization,
and the somewhat exotic duo of dielectronic recombination and resonant
excitation.  An even more obscure process called charge exchange (CX),
also known as charge transfer, is of interest to astrophysicists
chiefly for its role in determining the ionization balance
in H {\sc i} regions
via reactions such as 
$\mathrm{O}^{+} + \mathrm{H} \rightarrow \mathrm{O} + \mathrm{H}^{+}$
\cite{cit:field1971,cit:watson1976ism},
or in partially sweeping neutral H away from the Sun via
CX with solar wind protons, thus creating `pickup ions' and
contributing to the formation of a neutral H density enhancement
on the leading edge  of the heliosphere \cite{cit:blum1970}.

In the past decade, however, 
CX has gained the attention of X-ray astronomers
as the emission mechanism in a surprising range
of astrophysical sources, and for a variable background
present in {\em all} X-ray observations.
In Section 2 we review the basic features of charge exchange
emission and spectra, followed in Section 3 by astrophysical examples.
Section 4 discusses experimental methods
for studying CX, Section 5 presents results 
from electron beam ion trap experiments and compares them with theoretical
predictions, and Section 6 concludes with a discussion of
key issues to be addressed by
future laboratory and observational work.

\section{Charge Exchange Basics}
\label{sec:basics}

\subsection{The Mechanism}
\label{sec:basics-mech}

Charge exchange is the semi-resonant transfer of one or more
electrons from a neutral atom or molecule to an ion.  CX between
ions {\em can} occur, but the cross sections are very small because
of Coulombic repulsion.  By convention, the ion is referred to as
the `projectile' because it is usually moving faster than the
neutral `target.'  During the transfer collision, the energy
levels of both the projectile and target are distorted as the
electric fields of their nuclei and electron(s) are superposed.  At certain
internuclear distances, energy levels of the projectile and target
overlap at `curve crossings' so that an electron from the neutral target
may be radiationlessly transferred to the projectile ion.  
The cross section for this process is very large
compared to that for electron impact excitation,
typically of order $10^{-14}$ cm$^{2}$.

Because curve crossings occur at several internuclear distances,
electrons can be captured into several different energy levels,
imparting a semi-resonant character to the CX process.  If the recipient
ion is highly charged, it is the higher lying energy levels that
overlap with the neutral's energy levels.
A CX collision then results in the capture of an electron
into a high-$n$ level of the ion from which it can
radiatively decay, releasing an X-ray photon either
at the end of a radiative cascade or by decaying directly to the ground state.  

If the neutral species has more than one electron (i.e., is anything other
than atomic H), multiple electrons may be transferred, creating
a multiply excited ion that can either radiatively stabilize
or autoionize.
For the ions and collision energies
of interest here---those that give rise to X-ray emission---the cross
sections for multiple transfer are usually smaller 
than those for single-electron capture (SEC), and most of our discussion
will focus on SEC.

\subsection{General Equations}
\label{sec:basics-eqns}

It can be shown \cite{cit:janev1985} that for an ion of charge $q$,
the energy level having the maximum probability of being populated
in a collision with a H atom can be approximated by
\begin{equation}
n_{max} \sim q \left( 1 + \frac{q-1}{\sqrt{2q}} \right) ^{-1/2} \sim q^{3/4}.
\label{eq:nmax}
\end{equation}
For single-electron capture
from species other than H, $n_{max}$ can be approximated
by including an additional factor of
$\sqrt{I_{H}/I_n}$, where $I_{H}$ and $I_{n}$ are the ionization
potentials of H and the neutral target, respectively.

At relatively low collision energies (below $\sim$100 keV/amu),
the $n$ distribution is fairly narrowly peaked about
$n_{max}$.
As collision energy increases, the $n$ distribution
gradually broadens until a critical energy given by
\begin{equation}
	E_{crit} \sim 25 \sqrt{q} \mathrm{\;\;\; keV/amu}
\label{eq:ecrit}
\end{equation}
is reached.
At even higher energies,
$n_{max}$ slowly decreases and the distribution narrows again
\cite{cit:ryufuku1979}.

As an example, in the collision of O$^{8+}$ and H, the most likely
level to be populated in the resultant  O$^{7+}$ ion is
$n_{max}\sim5$.  At first glance this may seem rather surprising,
since the binding energy of that $n=5$ electron is 35 eV, much more
than the 13.6-eV binding energy of the H atom from which it was
captured.  
This is because
the electron sees a higher combined electric field as
the two nuclei approach each other; the curve crossing
occurs when the total binding energy of the electron
is significantly larger than its final binding energy
of 35 eV in the isolated O ion.  
In effect, some of the kinetic energy of the collision
is converted into potential energy, the so-called `energy defect.'
Following the electron capture, the O ion is left in an
excited state with an energy 836.5 eV above the ground state.
Details of the resulting decay, which releases at least one
photon (an X-ray), are given in Section~\ref{sec:results-hlike}.

\begin{figure}[t]
\centering
\includegraphics[angle=270,scale=1.00]{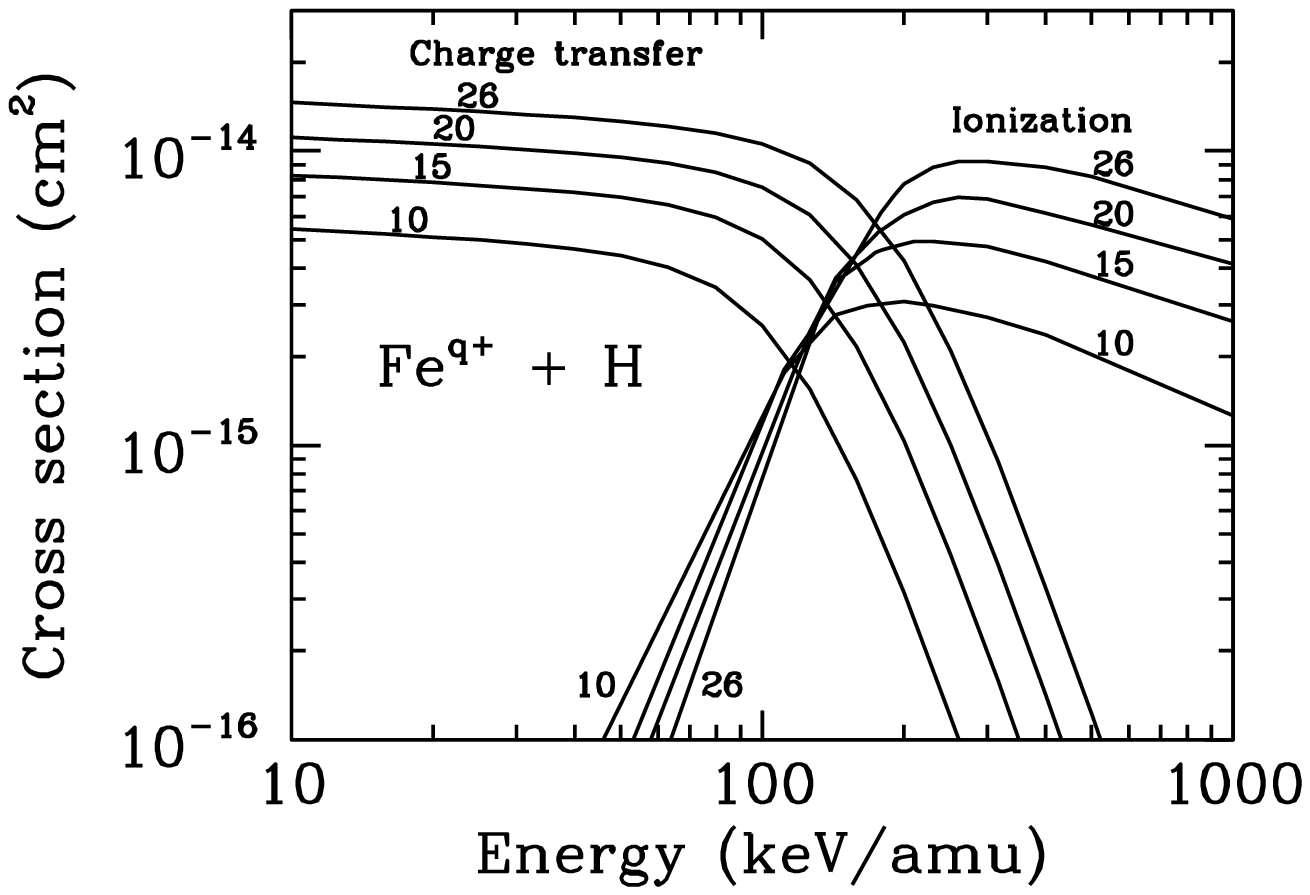}
\topcaption{
Theoretical cross sections 
for Fe$^{q+}$ $+$ H based on results from \protect\cite{cit:katsonis1991}.
Labels on each curve indicate the value of $q$.
\label{fig:kmjcross}
}
\end{figure}

Cross sections for CX have relatively little dependence on
the collision energy or target species.  At energies extending
from well below 1 keV/amu up to $E_{crit}$,
the cross section
can be roughly approximated as
\begin{equation}
\sigma_{CX} \sim q \times 10^{-15} \mathrm{cm^2}.
\label{eq:crudecross}
\end{equation}
At higher energies, the CX cross section rapidly declines while
the cross section for ionization of the projectile (known as
`charge stripping') rises and
then gradually falls (see Figure~\ref{fig:kmjcross}).


A more accurate approximation
for CX cross sections,
based on results from
the classical trajectory Monte Carlo (CTMC) modeling method
[see e.g., \citen{cit:abrines1966,cit:olson1981}],
with uncertainties of $\sim$30\% below $E_{crit}$
and roughly 50\% at higher energies
is provided by Katsonis, Maynard, and Janev
\cite{cit:katsonis1991}:
\begin{equation}
\tilde{\sigma} = \frac{A\ln{(B/\tilde{E})}}
        {1 + C\tilde{E}^{2} + D \tilde{E}^{4.5}} \; \; \;
        (10^{-16} \: \mathrm{cm}^{2})
\label{eq:kmjcross}
\end{equation}
where
$\tilde{\sigma} = \sigma_{CX}/q$,
$\tilde{E} = E/q^{0.5}$ (in keV/amu),
$A = 0.40096$,
$B = 2.44 \times 10^{6}$,
$C = 2.5592 \times 10^{-4}$,
and $D = 9.04985 \times 10^{-8}$.
There is no general equation for ionization cross sections below
$E_{crit}$, but $\sigma_{ioniz}$ peaks around $E=50/q^{0.415}$ keV
and scales at higher energies as $q^{2}/E$
\cite{cit:katsonis1991}.
At very low energies (up to a few hundred eV/amu) cross sections
may deviate significantly from Eqs.~\ref{eq:crudecross} 
and \ref{eq:kmjcross}
for some ion/target combinations.

\subsection{Collision Energy Regimes}
\label{sec:basics-Eranges}

Because CX cross sections have a relatively simple form when expressed
in terms of $E/q^{0.5}$
and $q$ is 
roughly one half 
the number of nucleons for
highly charged ions, collision energies are typically expressed in
terms of energy per atomic mass unit (e.g., eV/amu).  
The $\sqrt{q}$ dependence also appears in 
the equation for $E_{crit}$ (Equation~\ref{eq:ecrit}).
For ions of interest to X-ray astronomers --- mostly bare, 
H-like, and He-like ions from C to Fe, plus
L-shell ions of Fe and some lower-$Z$ elements) ---
$E_{crit}$ is roughly 100 keV/amu.
Energies below $E_{crit}$ are usually referred to as `low,'
energies above as `high,' and energies within a factor of two or three
of $E_{crit}$ as `intermediate.'
Below roughly 1 eV/amu, collisions are slow enough that
electron-electron interactions and pseudo-molecular effects
become important, and CX cross sections at these ultra-low energies
may differ significantly from those given by Equation~\ref{eq:kmjcross}.

The collision energies relevant to this paper are all above 1 eV/amu.
For comparison, 
typical temperatures of X-ray emitting plasmas range from roughly
$10^{6}$ to $10^{8}$ K.  From $(1/2)m v^2 = (3/2) k T$, a $10^{7}$ K
plasma temperature corresponds to 162 eV/amu for O ions and 46 eV/amu for Fe,
with collision energies 
scaling in proportion to $T$.
Although the fraction of neutral gas at such temperatures is
completely negligible for plasmas in equilibrium,
there are instances where astrophysical plasmas may undergo
CX with adjoining neutral gas clouds
(see Sections~\ref{sec:astro-pin} and \ref{sec:astro-gc}).

As discussed in Section~\ref{sec:astro-swcx}, 
the solar wind is responsible for the best known examples of 
X-ray CX emission
and can be divided into two main components: 
the slow wind, with a typical
velocity of 200--600 km/s (0.2--1.9 keV/amu),
and the fast wind, with a velocity of 600--800 km/s (1.9--3.3 keV/amu)
\cite{cit:smith2003}.
Note that the bulk-motion kinetic energy of solar wind ions is much larger
than the thermal energy acquired during
their formation in the solar corona, with
$T\sim2 \times 10^{6}$ K corresponding to $\sim$0.2 keV, or
of order 0.01 keV/amu.

Cosmic rays have much higher energies, greater than roughly 1 GeV/amu,
and their cross sections for CX are very small (see Figure~\ref{fig:kmjcross}).
There is, of course, a continuum of cosmic ray energies extending to
low energies, but the low-$E$ interstellar flux can not be directly
measured because of 'solar modulation,' in which
the solar wind and the magnetic field it carries act to deflect low energy
cosmic rays from penetrating the heliosphere \cite{cit:fulks1975}.
These particles, with energies from roughly 10 keV/amu to 1 GeV/amu
are sometimes referred to as suprathermal ions.
It has been
suggested that cosmic rays with energies below $\sim$1 MeV/amu
might contribute substantially to X-ray
line emission in the Galactic plane but, 
as discussed in Section~\ref{sec:astro-gc},
this is unlikely.

\subsection{Models and Multiple Electrons}
\label{sec:basics-models}

The preceding discussion and equations have focused on 
single-electron capture, which produces a singly excited
ion that always radiatively decays.
Multi-electron capture, of which double-electron capture (DEC)
is most common, opens up a much more complicated world
of autoexcitation, correlated double capture, true double capture,
correlated transfer and excitation, and autotransfer to Rydberg states.
Chesnel et al.\ \cite{cit:chesnel1999} present an overview of such processes
in the context of DEC in Ne$^{10+}$ $+$ He 
with astrophysically relevant collision energies
between 50 eV/amu and 15 keV/amu.

Modeling even SEC involving multi-electron targets
is much more difficult than for simple ion/H collisions,
and modeling DEC adds another large step of complexity.
The most common theoretical technique used in CX modeling is
the classical trajectory Monte Carlo (CTMC) method
\cite{cit:abrines1966,cit:olson1981},
which provides the predictions of level-specific population
distributions necessary for spectral modeling, unlike some other methods.
Although theoretical calculations for collisions involving
fully stripped ions and atomic H
are essentially exact because of the simplicity of the system
(just two nuclei and one electron), 
the CTMC method is difficult to extend to other systems
and most versions treat multi-electron targets simply as
atomic H with a modified ionization potential.
Not surprisingly, results from
experiments with multi-electron targets often
have significant disagreements with CTMC model predictions 
(see Section~\ref{sec:results-targets}).
Models also become more complicated as the number of projectile
electrons increases, and very little theoretical work has been done
on anything other than hydrogenic or fully stripped ions.

A discussion of current approaches 
to CX modeling is beyond the scope of this paper, but
Section 4 of the review on cometary CX emission by
Krasnopolsky, Greenwood, and Stancil \cite{cit:krasno2004}
gives an excellent summary of theoretical methods and results.
Each method has a limited range  of applicability, particularly in
terms of collision energy, but the limits are generally
not well known and model uncertainties are difficult
to estimate.  The problem is even greater when modeling
CX spectral emission, which depends very strongly
on the initial distribution of angular momentum states following
electron transfer (see Sections~\ref{sec:results-hlike} and
\ref{sec:results-helike}).  Experimental results are thus
vital in guiding the development of theory, but as 
seen in the Oak Ridge National Laboratory/University of Georgia
Charge Transfer Database for Astrophysics
({\tt http://cfadc.phy.ornl.gov/astro/ps/data/}),
very little experimental data
exist for astrophysically relevant CX collisions, even with H as the target.
Level-specific data (i.e., cross sections for individual $nl$ levels)
and spectra are even scarcer, often making it difficult to interpret
astrophysical observations, for which spectra are often
the only source of information.

\section{Astrophysical Relevance}
\label{sec:astro}


\subsection{Photoionized and Non-Equilibrium Plasmas}
\label{sec:astro-pin}

One area where the significance of CX has been appreciated for
some time is in calculations of ionization balance.
For collisional plasmas in equilibrium, CX is
generally of minor importance,
but in photoionized plasmas,
in interactions of the interstellar medium (ISM)
with stellar winds and cosmic rays,
and in non-equilibrium situations such as
supernovae and evaporating clouds,
CX can have a significant effect on the ionization balance, as well as
on the thermal evolution of the system via line emission.

As explained by several authors 
[e.g., \citen{cit:dalgarno1985,cit:ferland1997}],
CX can be at least as important as radiative recombination (RR) and
dielectronic recombination (DR) in photoionized plasmas;
the photoionization cross section of atomic hydrogen
decreases rapidly with photon energy ($\sigma_{PI} \propto E^{-3.5}$),
so highly charged ions can coexist with a non-negligible
population of neutral hydrogen and helium atoms.
Because cross sections for CX are usually orders of magnitude larger
than those for RR and DR, even a tiny fraction of neutral gas
can have a large effect on the overall plasma ionization state.

The tabulations of ionization and recombination cross sections
used in such calculations 
\cite{cit:arnaud1985,cit:arnaud1992,cit:mazzotta1998},
however, generally apply broad extrapolations
of the little data available, and use only
{\it total} CX cross sections.
If some emission lines are optically thick (e.g., in binary accretion disks),
the details of the CX spectrum become important;
for example, photons from optically thin high-$n$ transitions in H-like ions
(see Section~\ref{sec:results-hlike})
can escape from and thus cool dense gas much more easily than
Ly$\alpha$ photons.

CX may also play an observable role in the ionization (non)equilibrium of 
supernova remnants, particularly at the interfaces of hot shocked
plasma and relatively cool, dense clouds within the remnant.
Wise and Sarazin \cite{cit:wise1989} estimated how much CX X-ray emission
would result from a fast supernova remnant shock without such clouds
and concluded
that no more than about 10\% of the emission from He-like carbon
and nitrogen would arise from CX, and less from other elements.
Lallement \cite{cit:lallementICM2004} likewise calculated that
CX emission is minor for a supernova
blast wave propagating in a partially neutral ISM.
Wise and Sarazin suggested, however, 
that the role of CX would be enhanced by hydrodynamic
instabilities, or inhomogeneities in the interstellar gas
(particularly long filaments with large surface areas),
as may occur in relatively young remnants,
and Lallement noted that there are line-of-sight
enhancements to CX emission intensity along ionized/neutral boundaries.

Although there has been speculation over possible signatures
of X-ray CX in SNRs \cite{cit:rasmussen2001}, thus far
there is no clear evidence for such emission.
This may be in part because X-ray detector efficiency and
intrinsic resolving power are typically 
very low for the lines predicted to be most affected (He-like C and N),
making detection of any CX features difficult.

In addition to supernovae, Lallement \cite{cit:lallementICM2004} also 
considered CX emission from high velocity clouds moving through
the Galactic halo, and dense clouds interacting with galactic
winds or intra-cluster gas.  She concluded that CX might
contribute a significant fraction of the total emission in each case,
particularly in individual lines.
Detecting that emission will require
future large-area X-ray missions using non-dispersive imaging detectors
with good energy resolution, such as
the microcalorimeter \cite{cit:porter2004} 
used on the Lawrence Livermore Electron Beam
Ion Trap (see Section~\ref{sec:expt-ebit-magmode}).

\subsection{Solar Wind Charge Exchange}
\label{sec:astro-swcx}

\subsubsection{Comets}
\label{sec:astro-swcx-comets}

Renewed interest in astrophysical CX was triggered by
the discovery of X-ray emission from comets in a 1996 observation
of comet Hyakutake by the \rosat\ observatory \cite{cit:lisse1996}.
Soon afterward, Cravens \cite{cit:cravens1997} proposed that highly charged
heavy ions in the solar wind undergo CX with neutral gas
evaporating from the cometary nucleus
(mostly H$_{2}$O, CO, and CO$_{2}$).
Since then, more than twenty comets have been observed 
to emit X-rays and extreme ultraviolet radiation
by missions including \rosat, \euve, \asca, {\it BeppoSAX},
\chandra, \xmm, and \swift,
and the solar-wind charge exchange (SWCX) explanation is now
well established.

Cravens \cite{cit:cravens2002rev} reviewed cometary X-ray emission
in 2002, and a summary of observations up to 2004 is
provided by 
Lisse et al.\ \cite{cit:lisse2004}
and Krasnopolsky, Greenwood, and Stancil \cite{cit:krasno2004};
the latter also discuss CX modeling and laboratory experiments.
The most recent summary of comet observations (covering most of 2006, but not
Sasseen et al.\ \cite{cit:sasseen2006} or 
Krasnopolsky \cite{cit:krasnopolsky2006}) 
is part of a review by 
Bhardwaj et al.\ \cite{cit:bhardwaj2007}
that covers all sources of X-ray emission throughout the solar system.

\subsubsection{Spectra}
\label{sec:astro-swcx-spectra}

X-ray emission from comets was very surprising at first since
comets have temperatures of order 100 K while X-ray emission
is characteristic of million-degree matter.
Cravens \cite{cit:cravens1997} recognized that the
real source of the X-rays is the solar wind; 
highly charged ions in the wind capture electrons from neutral gas and
decay to ground, thus
releasing some of the potential energy of their
highly charged states that was originally acquired by thermal ionization in
the two-million-degree solar corona.

There are two main components of the solar wind:
the fast wind, with a typical velocity of 750 km/s,
and the slow wind, which is more highly ionized than the
fast wind and has an average velocity of 400 km/s
\cite{cit:smith2003}.
During solar minimum, when the Sun's corona is at its least active
point during the 11-year solar cycle (as it will be during 2007),
the slow wind is confined to within about 20 degrees of the
ecliptic while the fast wind is found at higher solar latitudes.
During solar maximum the two components are much more mixed,
with a predominance of slow wind at most latitudes.


\begin{table}[t]
\centering
\topcaption{Average abundances of Highly Charged Ions in the Solar Wind.
Element abundances (relative to O) are taken 
from Steiger et al.\ \protect\cite{cit:steiger2000}.
Ion fractions are from Schwadron and Cravens \protect\cite{cit:schwadron2000}.
\label{tab:swabunds}
}
\begin{tabular}{cccccc}
\hline
\hline
	& \multicolumn{2}{c}{Slow Wind}	&& \multicolumn{2}{c}{Fast Wind} \\
		\cline{2-3}			\cline{5-6}\\
	& Element	& Ion		&& Element	& Ion		\\
Ion	& Abund.	& Fraction	&& Abund.	& Fraction	\\
\hline
C$^{5+}$  & 0.67	& 0.210		&& 0.69		& 0.440		\\
C$^{6+}$  & 0.67	& 0.318		&& 0.69		& 0.085		\\
N$^{5+}$  & 0.078	& 0.065		&& 0.113	& 0.127		\\
N$^{6+}$  & 0.078	& 0.058		&& 0.113	& 0.011		\\
N$^{7+}$  & 0.078	& 0.006		&& 0.113	& 0.000		\\
O$^{6+}$  & 1.00	& 0.730		&& 1.00		& 0.970		\\
O$^{7+}$  & 1.00	& 0.200		&& 1.00		& 0.030		\\
O$^{8+}$  & 1.00	& 0.070		&& 1.00		& 0.000		\\
Ne$^{7+}$ & 0.097	& 0.004		&& 0.083	& 0.005		\\
Ne$^{8+}$ & 0.097	& 0.084		&& 0.083	& 0.102		\\
Mg$^{9+}$ & 0.145	& 0.052		&& 0.106	& 0.044		\\
Mg$^{10+}$ & 0.145	& 0.098		&& 0.106	& 0.029		\\
\hline
\end{tabular}
\end{table}


Although neutral target species may vary somewhat from one X-ray source
to another (e.g., H$_{2}$O in comets, H in planetary atmospheres), 
all SWCX spectra are similar because they share
the same source of emission---solar wind ions.
Table~\ref{tab:swabunds} lists average abundances of the `metals'
in the slow and fast solar wind that are most important to
X-ray CX emission \cite{cit:steiger2000,cit:schwadron2000}.
The most abundant metal ions are
the bare, H-like, and He-like charge states of C and O;
during coronal mass ejections, the fraction of highly charged
ions is much larger and may include bare and H-like Ne and Mg,
and enhancements in the population of L-shell Fe ions.
Bare ions charge exchange to produce H-like ions and Lyman series emission,
H-like ions charge exchange to produce He-like ions and what we will refer
to as `K-series' emission,
and He-like ions charge exchange to produce Li-like ions and $n \rightarrow 2$
emission.  (We focus on single-electron capture;
multi-electron CX is discussed in Section~\ref{sec:results-targets}).


\begin{figure}[]
\centering
\includegraphics[angle=270,scale=0.70]{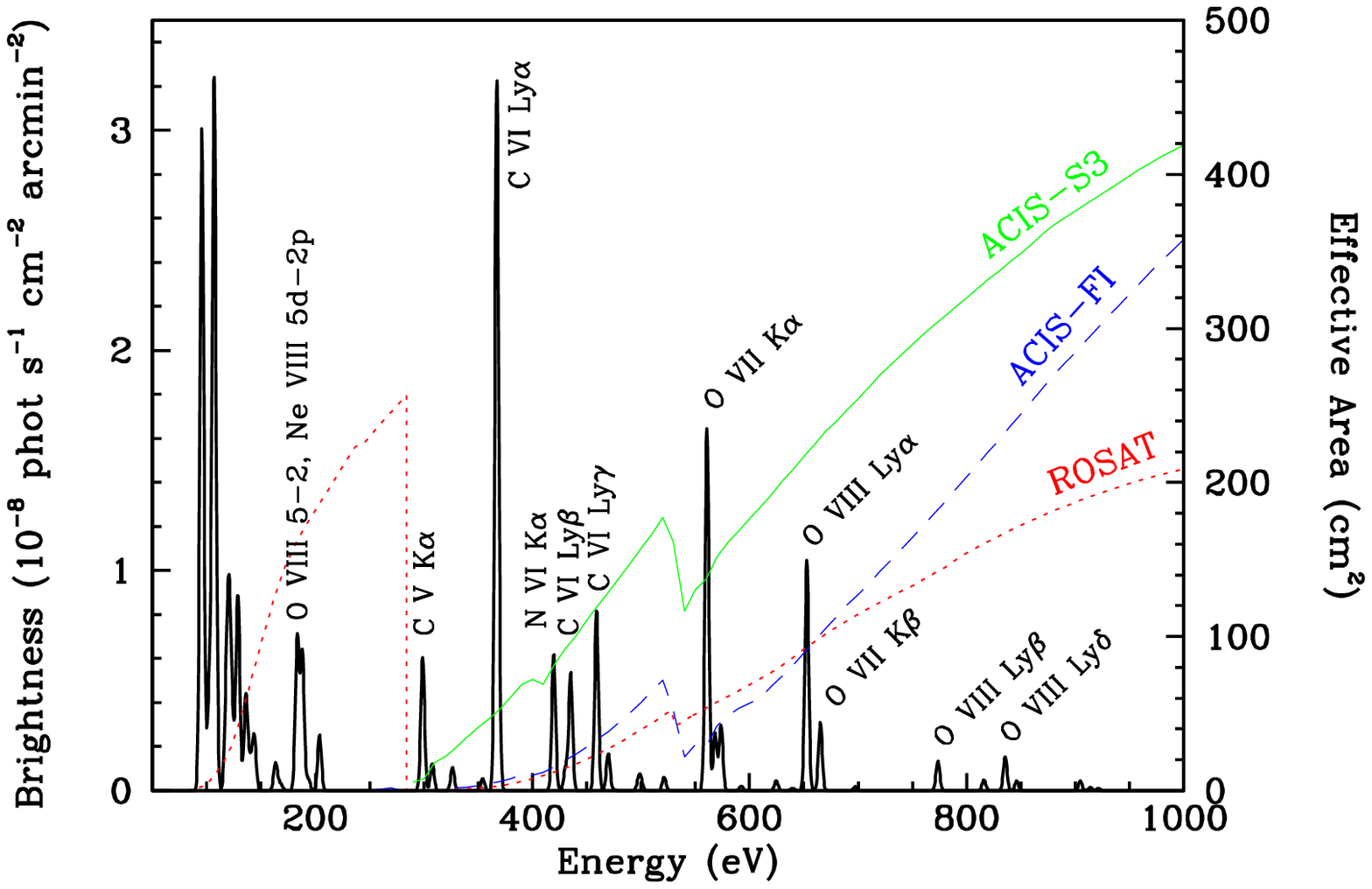}
\topcaption{Model geocoronal CX spectrum (4-eV resolution) for CX between 
slow solar-wind ions and atomic H, and effective areas for the
\rosat\ position-sensitive proportional counter and the
\chandra\ ACIS-S3 and front-illuminated (FI) CCDs.
Adapted from \protect\cite{cit:wargelin2004}.
\label{fig:modelSWCXspec}
}
\end{figure}

\begin{figure}[h]
\centering
\includegraphics[angle=0,scale=0.55]{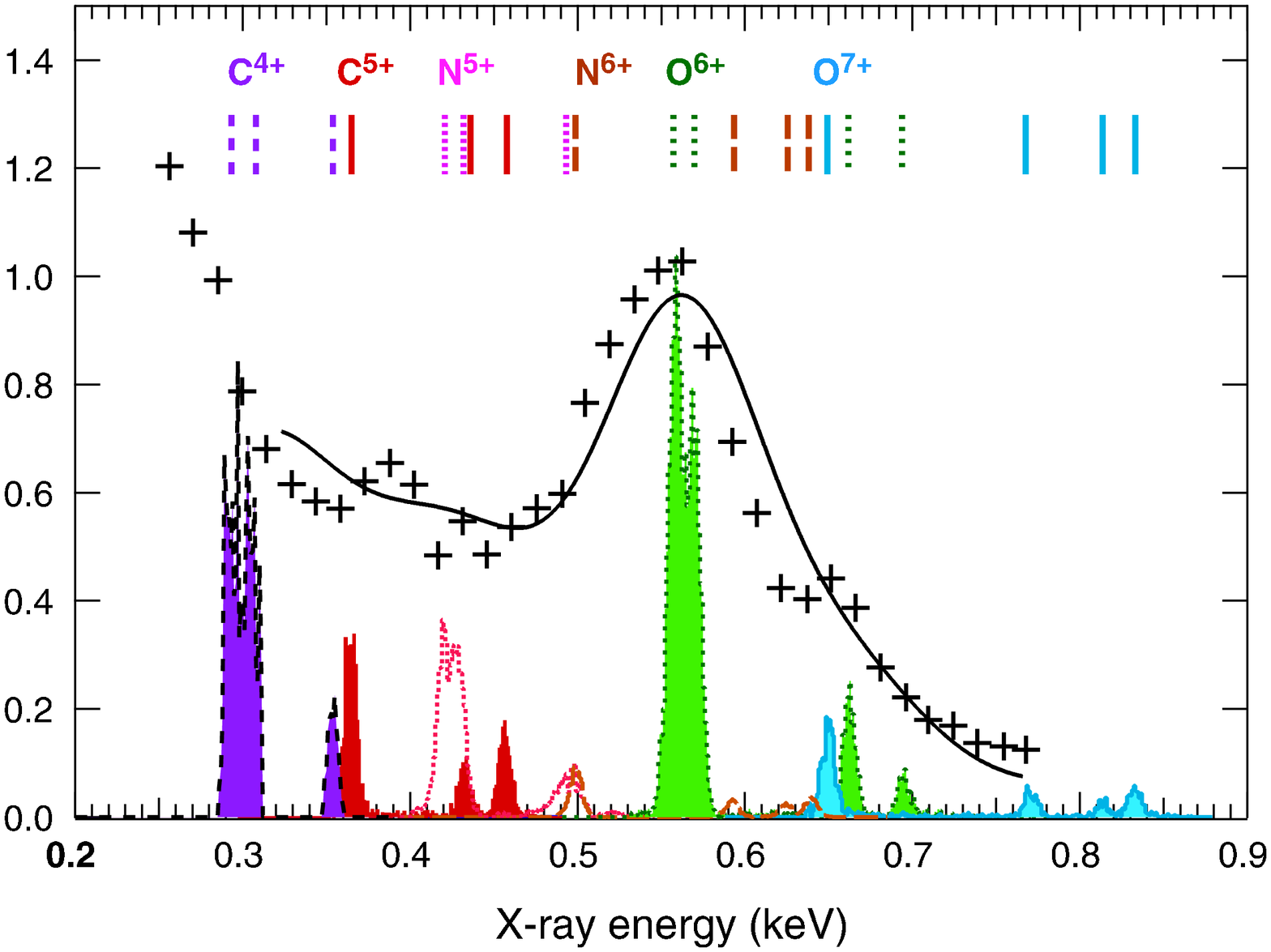}
\topcaption{Spectrum of Comet LINEAR C/1999 S4 observed by
\chandra\ ($+$), with fit ({\it solid line}) based on laboratory CX spectra 
obtained by the {\it ASTRO-E} microcalorimeter on the LLNL EBIT
(colored emission lines).
Adapted from \protect\cite{cit:beiersSci2003}.
\label{fig:cometspec}
}
\end{figure}

A number of theoretical SWCX spectra have been created
based on experimental cross sections
and calculated radiative rates, for CX between 
the slow or fast wind and H, He, or cometary neutrals
\cite{cit:kharchenko2000,
cit:krasno2002,
cit:rigazio2002,
cit:pepino2004}.
Model spectra for SWCX with H \cite{cit:wargelin2004}
are shown in Figure~\ref{fig:modelSWCXspec}.
Although carbon and oxygen have comparable SWCX X-ray intensities,
O emission lines are the most prominent features
in observed spectra because 
C lines are emitted at energies where X-ray detection efficiency is usually low.
An observed comet spectrum is shown in
Figure~\ref{fig:cometspec} along with a fit based on laboratory CX spectra
\cite{cit:beiersSci2003}.


\subsubsection{Planetary Atmospheres}

Soon after the discovery of cometary X-rays,
the concept of SWCX emission was expanded to include
emission from planetary atmospheres and the heliosphere.
\chandra\ and \xmm\ have each observed Mars 
\cite{cit:dennerl2002,cit:gunell2004,cit:dennerl2006},
revealing the presence of a faint halo of CX emission
surrounding the planet, in addition to fluorescent emission
from the illuminated disk.  The source of neutral gas for CX
is mostly atomic H in the extended outer atmosphere, or
exosphere, extending tens of thousands of km into space.
Although faint, \xmm\ grating spectra
clearly reveal CX emission lines of C and O at the expected wavelengths
\cite{cit:dennerl2006}.

CX emission has also been observed in Jupiter's aurorae
by \chandra\ \cite{cit:gladstone2002,cit:elsner2005}
and \xmm\ \cite{cit:branduardi2004,cit:branduardi2007},
although it is not clear if the emitting ions are from the solar
wind or the Io Plasma Torus
\cite{cit:cravensJup2003,cit:bunce2004,cit:kharJup2006}.  
In the former case, the CX emission
would comprise H-like and He-like C and O lines, while in the latter
the emission would be from O and L-shell S.
In either case, it appears the ions are accelerated by Jupiter's
magnetosphere to velocities of at least $\sim$5000 km/s ($\sim$130 keV/amu)
\cite{cit:branduardi2007}.
Particle measurements at Jupiter by the {\it New Horizons} mission in
February/March 2007 as it
receives a gravity boost on its way to Pluto,
in coordination with observations by \chandra, \xmm, {\it Hubble},
and other ground and space-based telescopes, should help resolve
many questions regarding Jupiter's auroral emission.

Venus has been observed multiple times by \chandra\ and
although its exospheric CX emission is expected to be stronger
than that of Mars because of its larger size and the higher
density of the solar wind,
no CX signal has yet been detected.
The first observation 
occurred in early 2001 near solar maximum when solar X-ray emission is at
its brightest, and fluorescent emission (O-K from atmospheric CO$_{2}$)
presumably obscured the predicted CX signal \cite{cit:dennerlVenus2002}.
Results from
subsequent observations conducted
in March 2006, near solar minimum, have not yet been published but
are expected to reveal CX emission
from O and perhaps C and Ne.  
Another \chandra\ observation is scheduled
for October 2007 in concert with {\it in situ} measurements of the solar wind
by {\it Venus Express}.

\subsubsection{Earth and Geocoronal CX}

CX emission from the Earth's exosphere has also been observed.
This geocoronal emission is present in the backgrounds
of all Earth-orbiting X-ray satellites, but its
presence is usually hard to discern.
Indications of its existence came during the \rosat\ mission, 
which conducted an all-sky survey
between 100 eV and 2.5 keV (usually broken into 1/4-keV, 3/4-keV,
and 1.5-keV bands).  Long-term enhancements (LTEs) in the observed
soft X-ray emission were occasionally observed \cite{cit:snowden1995},
and it was later noted that these enhancements
sometimes seemed to be correlated with 
solar wind variations \cite{cit:freyberg1998}.

\begin{figure}[h]
\centering
\includegraphics[angle=270,scale=0.55]{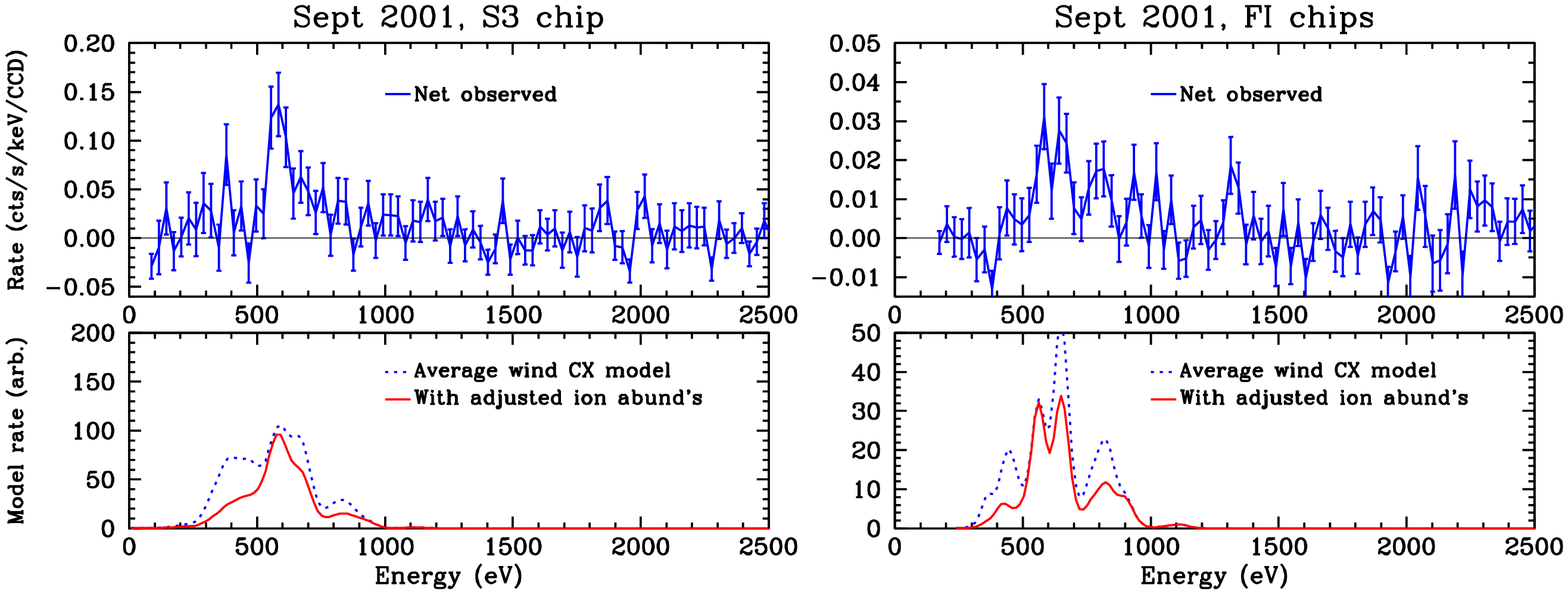}
\topcaption{
Top panels:
Background-subtracted spectra from 
\chandra\ observations of the dark Moon \cite{cit:wargelin2005},
showing geocoronal CX emission lines (see Figure 2).  
O \ka\ and \lya\ are nearly resolved in the FI spectrum,
which has better resolution than the S3 spectrum.
Bottom panels: Model CX spectra convolved with detector
responses; S3 has higher efficiency at low energies than
the FI CCDs. The dotted blue curve is for the average slow solar wind
(as in Figure 2).  The solid red curve has been adjusted to better
match the observed spectra by lowering
the carbon and H-like O abundances relative abundances.  
Mg is not included
in the model, but He-like Mg K$\alpha$ appears with $3\sigma$ significance in
the FI spectrum at 1354 eV.
Note that S3 and FI data were not collected entirely simultaneously.
\label{fig:darkmoon}
}
\end{figure}

\begin{figure}[h]
\centering
\includegraphics[angle=0,scale=0.43]{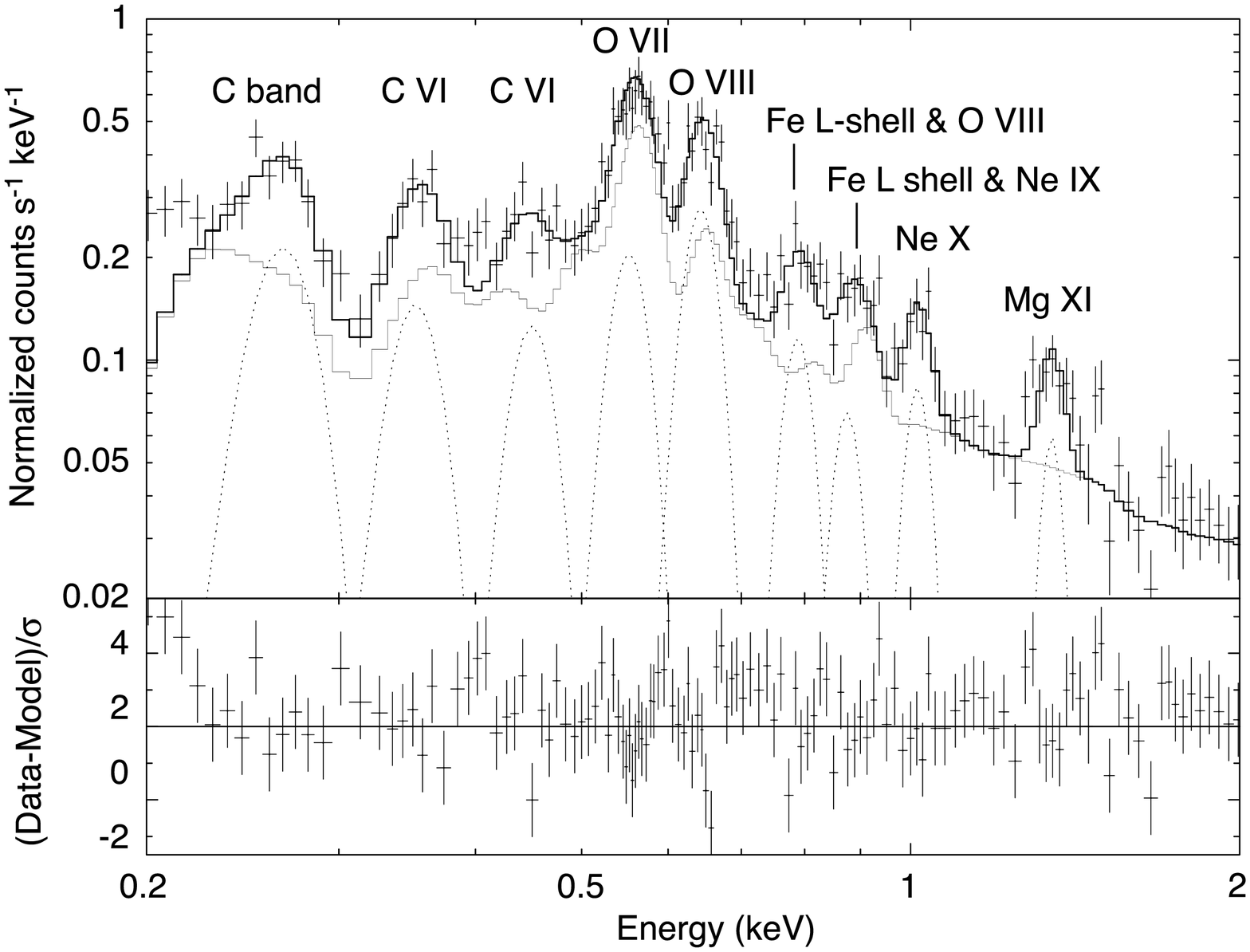}
\topcaption{
\suzaku\ observations comparing a baseline `stable' spectrum (solid thin line)
and a spectrum with enhanced emission (error bars) 
attributed to a transient
increase in the geocoronal CX X-ray background \cite{cit:fujimoto2007}.
The enhancement was fitted (solid thick) with nine lines (dotted) at
energies expected for CX emission.
\label{fig:xmmHDF}
}
\end{figure}

The temporal connection between solar wind variations, 
SWCX, and LTEs was
proven a decade after the \rosat\ survey
by Cravens, Robertson, and Snowden \cite{cit:cravensGeo2001},
and spectral confirmation was provided by a series of \chandra\ observations
of the Moon \cite{cit:wargelin2004}.
With the Moon blocking out the cosmic X-ray background,
the foreground geocoronal CX emission was revealed
(see Figure~\ref{fig:darkmoon})
with an intensity that approximately matched 
model predictions incorporating measured values of
solar wind density, velocity, element abundances, and ionization state.
Since then, LTE-like variations in the soft X-ray background 
have been noted in several observations by missions
including \chandra\ \cite{cit:mbm12,cit:hickox2006},
\xmm\ [\citen{cit:snowden2004,cit:bhardwaj2007} (Fig.~49)],
and \suzaku\ \cite{cit:fujimoto2007,cit:henley2007}.
An example is shown in 
Figure~\ref{fig:xmmHDF} comparing 
several \suzaku\ observations of the ecliptic pole \cite{cit:fujimoto2007}.

As discussed in papers modeling the interaction of
the solar wind with the Earth's magnetosphere
\cite{cit:robertson2003JGR,cit:robertson2003GRL,cit:robertson2006},
geocoronal emission is strongest in the sunward direction
where the solar wind compresses the magnetosphere,
allowing solar wind ions to charge exchange in relatively dense
regions of the outer atmosphere.  In the downwind direction,
geocoronal emission is negligible because
the solar wind is excluded from the magnetotail.
The SWCX emission observed by X-ray satellites depends not
only on where they are looking, but also where they 
are looking from, which may be well above or within
regions of high CX emissivity.

%

%

\subsubsection{The Heliosphere}

Although the density of neutral gas in the heliosphere is much less
than in the geocorona (roughly 0.1 cm$^{-3}$ versus of order 10 cm$^{-3}$),
the volume and total amount of gas is much larger.  Soon after cometary
X-ray emission was recognized to be the result of SWCX, 
Cox \cite{cit:cox1998} suggested that
a diffuse glow of CX emission might permeate the entire heliosphere
as neutral H and He from the interstellar medium flowed into
the solar system and interacted with the solar wind.
Cravens \cite{cit:cravensHelio2000} studied this idea quantitatively and
concluded that heliospheric CX emission (as viewed from Earth)
was roughly ten times as strong as geocoronal X-ray emission
and might account for up to half of the soft X-ray background.
%
%
Subsequent theoretical work 
\cite{cit:cravensGeo2001,
cit:robertson2003JGR,
cit:pepino2004,
cit:lallement2004,
cit:wargelin2004,
cit:koutroumpa2006,
cit:robertson2006}
has investigated how
heliospheric X-ray emission intensity varies with
observer location, look direction, the solar cycle, and 
temporal variations in the solar wind.
Estimates of the SWCX contribution to the soft X-ray background
range from as little as 5--10\% in
some directions to nearly all the background in some energy bands,
although model uncertainties are large.
Observations of stellar wind CX have also been proposed 
to image astrospheres and
measure mass-loss rates around nearby late-type stars \cite{cit:wargelin2001}
but existing X-ray observatories do not have quite enough
sensitivity \cite{cit:wargelin2002}.

Observational progress on geocoronal and heliospheric CX
\cite{cit:mbm12,cit:snowden2004,cit:fujimoto2007,cit:wargelin2004,
cit:smith2007,cit:hurwitz2005,cit:henley2007}
has been slower than for theoretical work
given the difficulty of distinguishing CX emission from 
the true cosmic X-ray background and the weakness
of the CX signal, which is typically of the same order as 
detector background.
Microcalorimeter X-ray detectors, which have much better energy
resolution ($\la 6$ eV FWHM) than the CCDs on \suzaku, \xmm,
and \chandra\ (roughly 50--100 eV FWHM below 1 keV), hold great promise
for studies of astrophysical CX, 
as they can easily discern spectral
features that uniquely identify CX emission 
(see Sections~\ref{sec:results-hlike} and \ref{sec:results-helike}).

In addition to its potential as a means to study the Earth's
magnetosphere, the heliosphere, and stellar winds and astrospheres
around other stars, the existence of SWCX impacts many other
astrophysical studies.  Our understanding of the local interstellar
medium is obviously affected; the temperature and pressure
of the Local Bubble are likely lower than has been previously assumed
\cite{cit:lallement2004}, and SWCX may explain the mismatch between
Local Bubble emission measures, metal depletions, and H column
densities derived from extreme ultraviolet versus X-ray observations
\cite{cit:hurwitz2005,cit:smith2007}.
Uncertainty in the SWCX level also affects spectral analyses of
extended objects such as supernova remnants and galaxy clusters
(e.g., apparently redshifted O {\sc vii} K$\alpha$ cluster emission
can be explained by local SWCX emission 
\cite{cit:bregman2006,cit:nevalainen2007}),
and is the limiting factor in efforts
to detect emission from the Warm-Hot Intergalactic Medium.




\subsection{Galactic Center and Galactic Ridge}
\label{sec:astro-gc}

While SWCX is now well established, the origin of the Galactic Ridge
X-ray Emission (GRXE) remains controversial a quarter
century after its discovery \cite{cit:worrall1982}.
The Galactic Ridge lies within a few degrees of the Galactic Plane
and within roughly 45 degrees of the Galactic Center (GC), which
it includes.
It features apparently diffuse X-ray emission with a spectrum
that is virtually the same all along the Ridge, apart from intensity.
In addition to strong continuum radiation there are
prominent He-like and H-like emission lines from Mg, Si, S, Ar, Ca, and Fe
\cite{cit:koyama1996,cit:kaneda1997,cit:ebisawa2001,cit:muno2004}.
Several explanations for the line emission have been proposed
[\citen{cit:muno2004,cit:revniv2006} and references therein],
including the
idea that it arises from the CX of low energy cosmic rays 
with neutral gas in the Galactic Plane \cite{cit:tanaka1999,cit:tanaka2002}.
An earlier version of this idea was first suggested by Silk and Steigman
\cite{cit:silk1969}
but subsequent theoretical studies 
\cite{cit:watson1976,cit:bussard1978,cit:rule1979}
concluded that
the fraction of nearly fully ionized cosmic rays needed to produce
the observed H-like and He-like line emission was negligible
below a few MeV/amu (well above $E_{crit}$), 
and the cross sections for CX at 
higher energies were much too small to explain the observed GRXE intensities.

The most recent and likely explanation is a revival of an earlier idea.
Based on extrapolations of {\it Rossi-XTE} data and noting the close similarity
between the GRXE and infrared background emission distributions,
Revnivtsev et al.\ \cite{cit:revniv2006}
propose
that the GRXE arises from the superposition
of unresolved coronally active binaries and cataclysmic variables.
Other authors \cite{cit:ebisawa2001,cit:muno2004}, however, argue that the 
required sources would have
been resolved in existing \chandra\ observations of the Galactic Center
and that the GRXE is truly diffuse.  
Spectral comparisons of the GRXE with Fe line diffuse emission from the center 
of the nearby starburst galaxy M32 are inconclusive \cite{cit:strickland2007}.

Although the cosmic-ray CX hypothesis
looks unlikely, it might be revived if some mechanism is found to
extend the range of nearly fully stripped cosmic rays 
to energies below 1 MeV/amu.
(Recall from Section~\ref{sec:basics-Eranges} 
that no unbiased measurements of interstellar
cosmic rays exist below about 1 GeV/amu.)  Possibilities may include larger
than assumed rates of multi-electron ionization 
of cosmic rays in sub-MeV/amu collisions \cite{cit:rule1979}
(for which there are very little experimental data)
or reionization of cosmic rays by Galactic
magnetic field reconnection \cite{cit:kronberg2004}.
The theoretical studies from the 1970's
\cite{cit:watson1976,cit:bussard1978,cit:rule1979}
also overlooked increased metal abundances closer to the GC.

The Galactic Center is the brightest part of the Galactic Ridge,
with intense thermal
emission from abundant hot gas.
There are also a number of warm dense molecular clouds, so
CX emission from the thermal highly charged ions must be present
at some level.  However, diffuse thermal emission together with
radiation from the dense population of point sources in the GC
is very likely to obscure any signature of CX emission,
whether from cosmic rays or the interfaces of
hot gas and warm clouds.
This is confirmed in recent observations of 
the Galactic Center region by \suzaku\ that indicate
the centroid of the He-like Fe \ka\ complex has an
energy too high to be explained by CX alone
\cite{cit:koyama2007,cit:wargelin2005} (see Section~\ref{sec:results-helike}).

Although more in the realm of gamma-rays than X-rays,
solid evidence for the operation of CX in the GC
comes from the measurement of the 511-keV annihilation line by
INTEGRAL.  Based on the line width and shape,
Churazov et al.\ \cite{cit:churazov2005}
and Jean et al.\ \cite{cit:jean2006}
estimate that nearly all of the observed emission comes from the
annihilation of positronium 
(as opposed to direct $e^{+}$ and $e^{-}$ collisions) 
and that roughly half of the positronium is
formed via the CX of positrons with neutral gas.
The positron and electron eventually annihilate, releasing
either two 511-keV photons if the $e^{+}$ and $e^{-}$ have
opposite spins, or three lower-energy photons with a continuum
of energies if the spins are the same.

\section{Experimental Methods}
\label{sec:expt}

Until the development of electron beam ion traps (EBITs), virtually 
all measurements of charge exchange involved crossed-beam experiments 
in which a beam of highly charged ions passes through a source 
of neutral gas, usually a supersonic gas jet (or `gas cell') inside a chamber 
with differential pumping. 
Ion beams can be generated by foil stripping and then injected 
into a storage ring and held at a constant (very high, MeV/amu) 
energy (e.g., \cite{ma01}). As they circle the ring, the ions repeatedly 
enter a gas cell where they can undergo CX.  
Electron cyclotron resonance (ECR) sources and 
electron beam ion sources (EBISs) are also used to provide ions 
in crossed-beam experiments, typically at energies of a few or tens 
of keV/amu.  Measurements in crossed-beam experiments usually focus 
on determinations of ion charge, which yield CX cross sections.
Studies of photon emission are rare, but are becoming more common.

Some of the first charge-exchange-induced emission spectra of 
K-shell ions were obtained in a crossed-beam experiment by 
Greenwood et al.\ \cite{greenwood00,greenwood01}. They
studied bare Ne$^{10+}$ and several other lower-$q$ ions 
from an ECR 
source at the Jet Propulsion Laboratory interacting with 
H$_2$O, He, H$_2$, and CO$_2$. The data were obtained at ion-neutral 
collision energies of $\sim$3 keV/amu, comparable to ion energies
in the fast solar wind.
Similar measurements conducted with the ECR source at the 
University of Nevada, Reno, demonstrated the role of multiple 
electron capture in x-ray line formation by CX \cite{ali05}. 
One complication common to all merged or crossed-beam experiments 
is that ions in metastable states (such as the $1s2s\:^1S_3$ state 
in He-like ions) may move well beyond the interaction region before 
they radiatively decay, making spectral measurements incomplete or 
harder to interpret. 
Nevertheless,
the combination of charge-state determinations and spectral measurements 
is a very powerful one, potentially revealing state-specific 
absolute cross sections for both single- and multi-electron CX. 
Some relatively recent enhancements of the cross-beam technique
provide additional information, particularly state-selective
cross sections (i.e., cross sections for individual $n$ and sometimes $l$
levels) by measuring scattering angles and/or ion kinetic energy losses
(`$Q$ values'; e.g., 
\cite{cit:chesnel1999,ali1994,hasan2001,flechard2001,fischer2002,edgufry2004}).

\begin{figure}
\centering
\includegraphics[angle=0,scale=0.70]{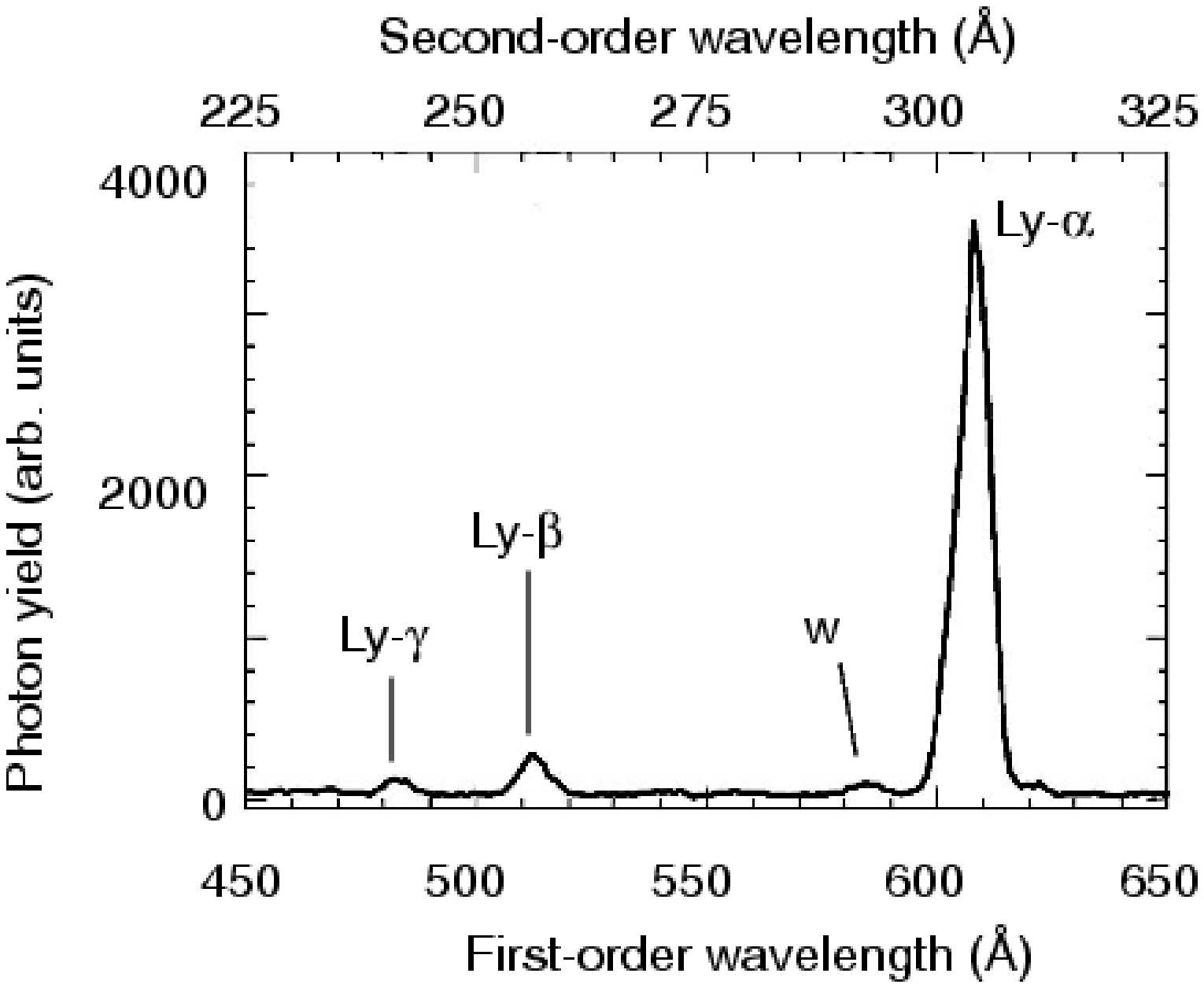}
\topcaption{
Spectrum of the Lyman series of He$^+$ formed by CX of 
He$^{2+}$ with H$_2$O measured in second order of reflection. 
Also seen is the $1s2p\:^1P_1 \rightarrow 1s^2\:^1S_0$ emission ($w$) 
in neutral helium. The latter transition is observed in first order 
and is attributed to double-electron capture during CX. 
The spectrum was obtained at the 
Kernfysisch Versneller Instituut in Groningen \cite{bodewits05,seredyuk05} 
at a collision energy of 48 keV.
\label{fig:HeCXspec}
}
\end{figure}

One example of a K-shell ion that has been studied using
the crossed-beam method 
is He$^{2+}$.
The ionization potential of hydrogenic He$^{1+}$ is 54.4 eV (corresponding
to $\lambda=228$ \AA),
so CX reactions involving bare helium form lines in the 
extreme ultraviolet. A spectrum from CX with H$_2$O is shown in 
Figure~\ref{fig:HeCXspec}.
The measurement was carried out using 50 keV He$^{2+}$ ions at the 
Kernfysisch Versneller Instituut in Groningen \cite{bodewits05,seredyuk05}. 
While most of the emission is the result of single-electron capture, 
a small feature corresponding to the $1s2p\:^1P_1 \rightarrow 1s^2\:^1S_0$ 
transition in neutral He is seen. This feature is produced only by 
``true'' double-electron capture, where ``true'' means that 
both of the transferred electrons radiatively stabilize,
as opposed to one of them being lost by autoionization.
As shown by Lubinski et al.\ \cite{lubinski01}, the cross section 
for single capture producing the Ly$\alpha$ line at 304 \AA\ drops 
as the collision energy decreases, while that for producing the 
neutral helium line, labeled $w$ at 584 \AA\ in 
Figure~\ref{fig:HeCXspec}, increases. 
As a result, the ratio of the Ly$\alpha$ line 
in He$^+$ to that of $w$ in neutral helium is dependent on the 
collision energy and can be used to determine the velocity of 
alpha particles in solar wind interacting with 
cometary neutrals \cite{bodewits04}.


\begin{figure}
\centering
\includegraphics[angle=0,scale=0.50]{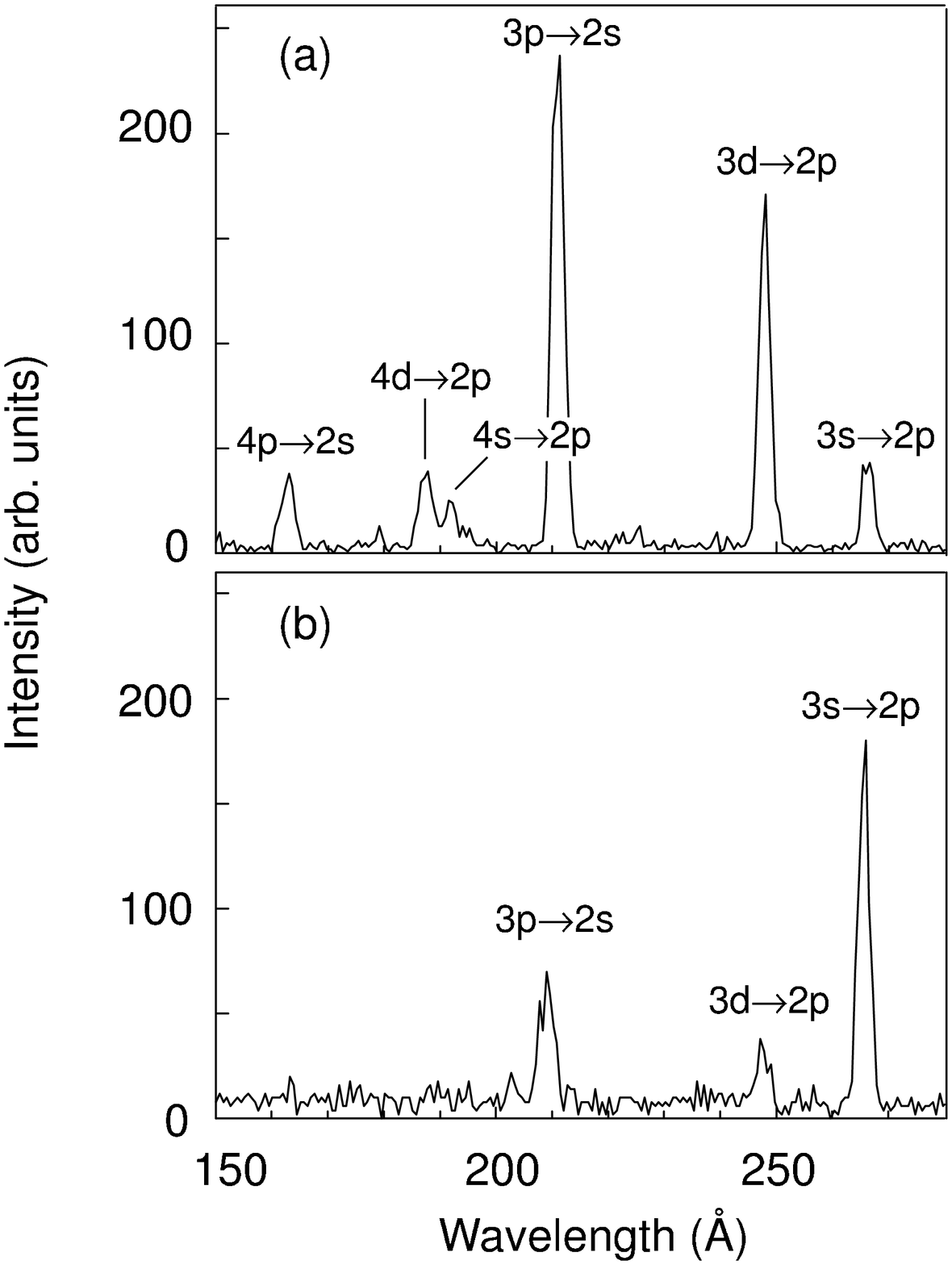}
\topcaption{
EUV spectrum of N$^{4+}$ formed by CX of N$^{5+}$ with
H$_2$. The spectra were recorded using the 
crossed-beam facility in Groningen by \cite{lubinski00} 
at collision energies of (a) 16.5 keV and (b) 0.5 keV.
\label{fig:LshellN}
}
\end{figure}

Crossed-beam experiments have also been used to determine 
CX cross sections and spectral emission from L-shell ions. 
Such measurements have concentrated on relatively simple L-shell ions, 
e.g., Li-like or Be-like ions \cite{bliek98,lubinski00,ehrenreich05}, 
which are close analogs of H-like and He-like ions, respectively. 
However, additional lines appear in these spectra because of the 
fact that the $n=2$ ground configuration may assume both an 
$\ell=0$ and $\ell=1$ angular momentum state. This is illustrated in 
Figure~\ref{fig:LshellN}, which shows the L-shell emission of 
Li-like N$^{4+}$ formed in the interaction of N$^{5+}$ with H$_2$ 
using the crossed-beam facility at Groningen \cite{lubinski00}.
Note that the shape of the L-shell 
emission depends on the collision energy, similar  to the
behavior of H-like spectra (see Section~\ref{sec:results-hlike}).


Something like the inverse of the crossed-beam setup takes place 
in tokamaks, where a neutral beam (usually of H isotopes) in injected 
into the tokamak's plasma. Trace concentrations of highly ionized elements, 
either deliberately introduced or present as background impurities, then 
charge exchange with the neutrals in the beam. 
The collision energy is largely determined by the neutral target energy, 
typically tens of keV/amu, while the thermalized ions have energies 
of order 100 eV/amu.  Using this approach, a spectrum of 
He-like Ar$^{16+}$ formed by CX of thermal Ar$^{17+}$ ions 
with an 80 keV beam of neutral deuterium was 
recorded on the National Spherical Tokamak Experiment (NSTX) at 
Princeton \cite{beiers05b}.

In tokamaks there is a continuous influx of neutral H gas into the plama 
because of hydrogen recycling at the wall and the plasma limiters.  
Deep inside the plasma, little neutral hydrogen exists and signal 
produced by CX competes with electron-impact excitation so that the 
contribution from CX is nearly impossible to isolate. 
However, tokamaks exhibit a low level of emission arising from CX 
with trace amounts of neutral H gas near the plasma boundary. 
In fact, X-ray emission produced by CX near the plasma edge has 
been observed even from highly charged ions \cite{kallne84ax,rice86} 
as ion transport diffuses highly charged ions to the edge region where 
they can interact with neutral hydrogen.

\subsection{Electron Beam Ion Traps}
\label{sec:expt-ebit}

This review focuses on CX results from EBITs, which differ from 
crossed-beam experiments in that the emitting ions are more or less 
stationary so that complete 
spectra can be obtained, even from long-lived metastable states. 
Unlike merged beams, however, EBITs cannot provide absolute cross sections 
because necessary experimental parameters such as neutral gas density 
can not be determined with enough precision. 

\begin{figure}[t]
  \centering
  \begin{minipage}[t]{0.48\textwidth}
    \centering
    \includegraphics[angle=90,width=0.95\textwidth]{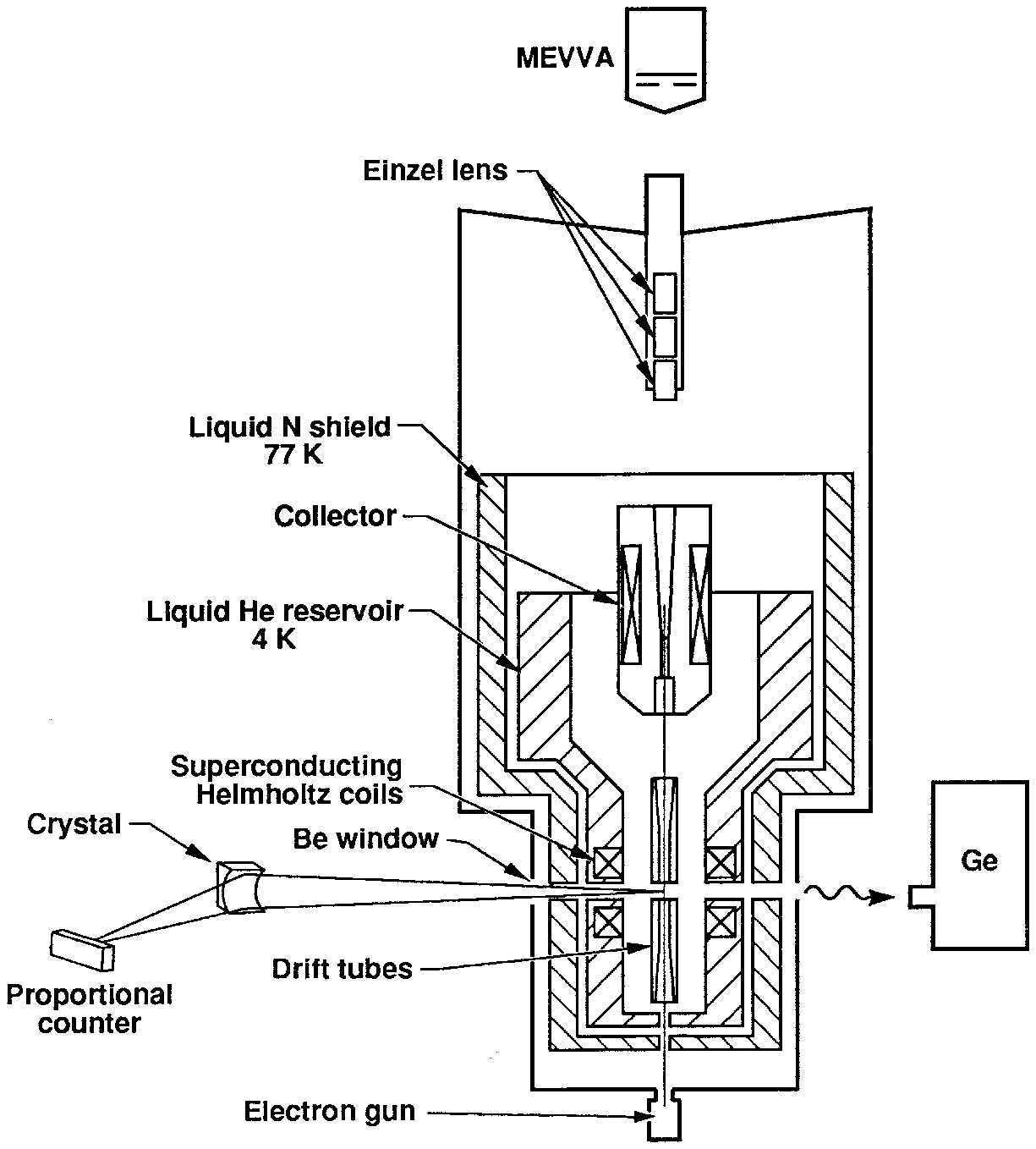}
  \topcaption{
Schematic cross section of the Livermore EBIT.
Highly charged ions are created by electron collisions
and confined in the trap region.
Various X-ray detectors ring the trap region to
observe the emitted photons.
  \label{fig:ebitcross}
        }
  \end{minipage}
  \hfill
  \begin{minipage}[t]{0.48\textwidth}
    \centering
    \includegraphics[angle=90,width=0.95\textwidth]{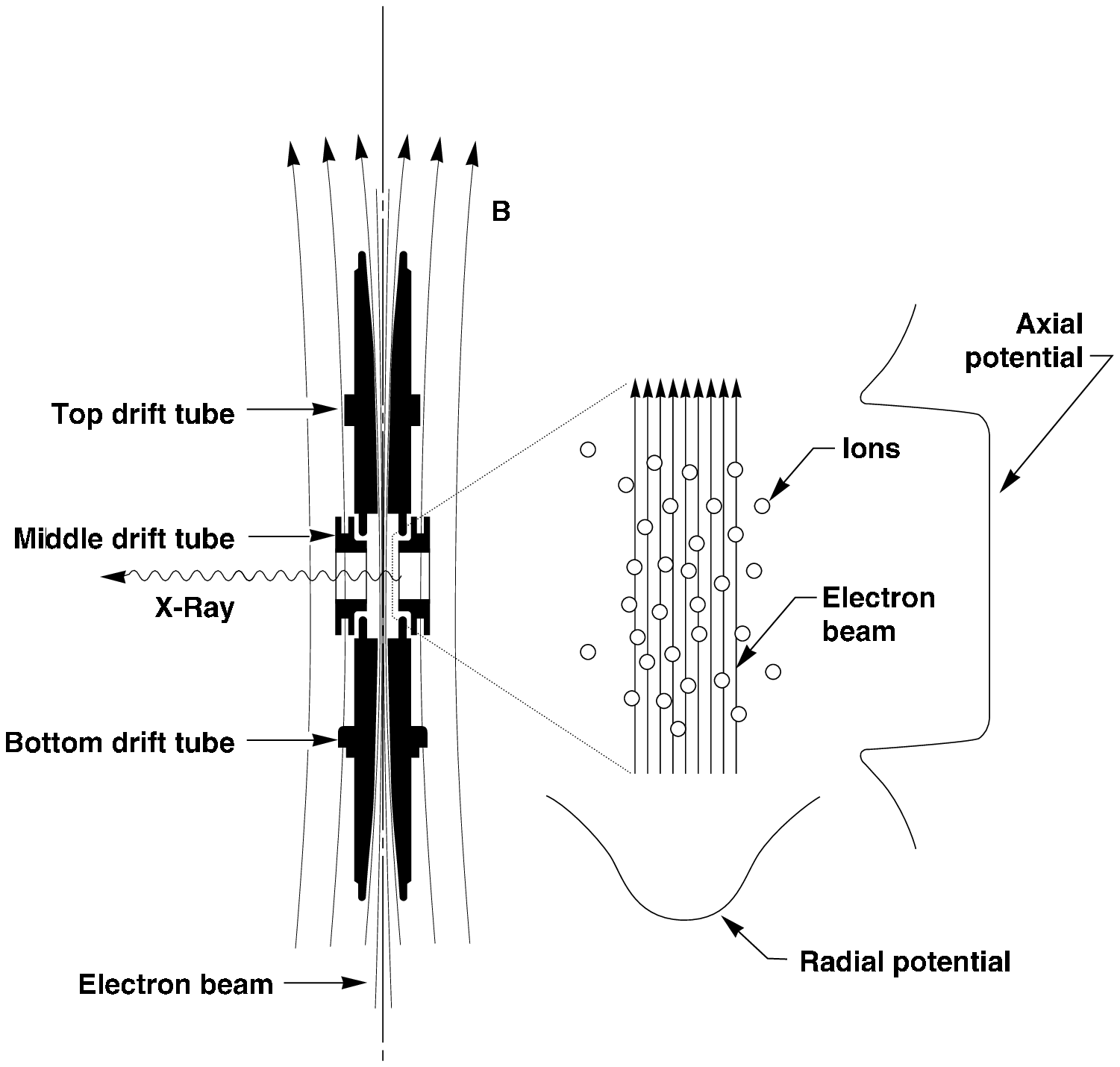}
  \topcaption{
EBIT drift tube assembly, showing the three
drift tubes and the central trap region.
The axial potential well is generated by raising
the voltage of the two end drift tubes above that of the middle.
Ions are radially confined by the electrostatic attraction
of the electron beam, which also collisionally ionizes the ions.
In the magnetic trapping mode the electron beam is off and
radially confinement is provided by the 3-Tesla magnetic field.
\label{fig:ebittrap}
}
  \end{minipage}
\end{figure}

The design and operation of EBITs is described in this volume 
(R.E.~Marrs, this issue) and elsewhere \cite{cit:levine1988,cit:levine1989}.
EBITs are based 
on electron beam ion sources but modified specifically 
for spectroscopic studies of the interaction of highly charged ions 
with an electron beam.  The electrons pass through a 2-cm-long trap 
region and are focused by a $\sim$3-Tesla magnetic field into a 
beam about 60 $\mu$m in diameter (see Figures~\ref{fig:ebitcross}
and \ref{fig:ebittrap}).  
Neutral atoms or ions with low charge from a vacuum arc source are injected 
into the trap where they are collisionally ionized and excited by the 
electron beam, which can be tuned to energies between 
a few hundred eV and tens or even hundreds of keV. 
Very high charge states can be attained, with longitudinal 
electrostatic confinement of the ions provided by a set of drift tubes. 
Radial confinement is provided by electrostatic attraction of 
the electron beam, as well as by the magnetic field. 
Several slots cut into the drift tubes are aligned with vacuum
ports, providing access for spectrometers and other instrumentation.

CX is both a blessing and a curse in EBITs. Early experiments on 
the original LLNL EBIT were unable to reach the 
ion charge states expected for a given electron beam energy because the 
most highly charged ions acquired so much kinetic energy 
in collisions with electrons 
that they escaped from the trap. 
It was noticed that X-ray emission rates, indicative of the number of 
trapped high-charge ions, would increase momentarily after occasional 
high-voltage breakdowns, which suggested that a lower-quality vacuum 
caused by the release of contaminants on internal surfaces during the 
electrical shorts somehow improved trap performance (R.E.\ Marrs,
private comm.).

The solution was to intentionally inject a neutral low-$Z$ gas 
(usually N$_{2}$, although neutral Ti evaporated from a hot wire was also
used early on), providing light ions 
that collided with the heavier ions of interest 
and carried away excess kinetic energy by escaping from the trap 
in a process called evaporative cooling 
\cite{cit:penetrante1989,cit:schneider1989,cit:penetrante1991}.
Because the energy required for escape is proportional to $qV_{trap}$, 
lower-$Z$, lower-$q$ coolant ions escape first and 
allow higher charge states to be attained for the ions of interest. 
In competition with this process is CX between 
neutral gas, either ambient or deliberately injected, and the 
ions under study.  Too much neutral gas suppresses the 
overall charge balance, and even tiny leaks in EBIT 
internals can substantially degrade performance. 

The collision energy in a CX reaction is given by the energy of the ion, 
as the energy of the neutral is negligible by comparison. Ion energies are 
primarily determined by ion charge and trap depth, i.e., by the fact that 
ions with energy higher than $qV_{trap}$ leave the trap.  Ion temperatures
have been determined for several combinations of ion charge and trap depth 
by measuring Doppler-broadened widths of selected x-ray lines with 
very high resolution crystal spectrometers 
\cite{cit:doppler1,cit:doppler2,cit:doppler3,cit:doppler4}.
Trapped ions were found to have energies below the
the $qV_{trap}$ limit, which is not surprising because they interact with 
lower-$Z$ ions used for evaporative cooling.  Most experiments run with 
characteristic ion energies between 1 and 20 eV/amu.

\subsubsection{Magnetic Trapping Mode}
\label{sec:expt-ebit-magmode}

In normal operation, with the electron beam turned on, 
i.e., in the `electron trapping mode,' 
electron-ion collisions ionize and excite the trapped ions. 
In `magnetic trapping mode' \cite{cit:beiersMag1996}
the electron beam is turned off.  Longitudinal trapping by the
potential well is unaffected 
and the ions remain confined in the radial direction, although more loosely, 
by the strong magnetic field. 

The main applications of the magnetic trapping mode have been 
in measurements of radiative lifetimes and CX \cite{schweikhard02}. 
In both cases, the emission lines of interest are revealed once 
emission from electron impact excitation 
(also known as direct excitation, or DE) ceases. 
To illustrate the difference in DE and CX emission rates, consider
\lya\ emission from direct excitation of H-like Ne$^{9+}$ and
from CX of fully stripped Ne$^{10+}$ with atomic H.
The emission rate for DE (in cts/s) is
\begin{equation}
R_{DE}=N_{i} n_{e} \sigma_{DE} v_{e}
\end{equation}
where
$N_{i}$  is the number of emitting ions,
$n_{e}$ is the electron density,
$\sigma_{DE}$ is the cross section for direct excitation of \lya,
and $v_{e}$ is electron velocity (effectively the collision velocity
because ion velocity is much smaller).
The corresponding equation for CX (using the approximation
that every CX reaction produces a \lya\ photon) is
\begin{equation}
R_{CX}=N_{i} n_{n} \sigma_{CX} v_{i}
\end{equation}
where
$n_{n}$ is the density of neutral gas,
$\sigma_{CX}$ is the CX cross section for Ne$^{10+}$,
and $v_{i}$ is ion velocity (effectively the collision velocity
because the neutral gas velocity is much smaller).

\begin{figure}[h]
\centering
\includegraphics[angle=0,scale=0.50]{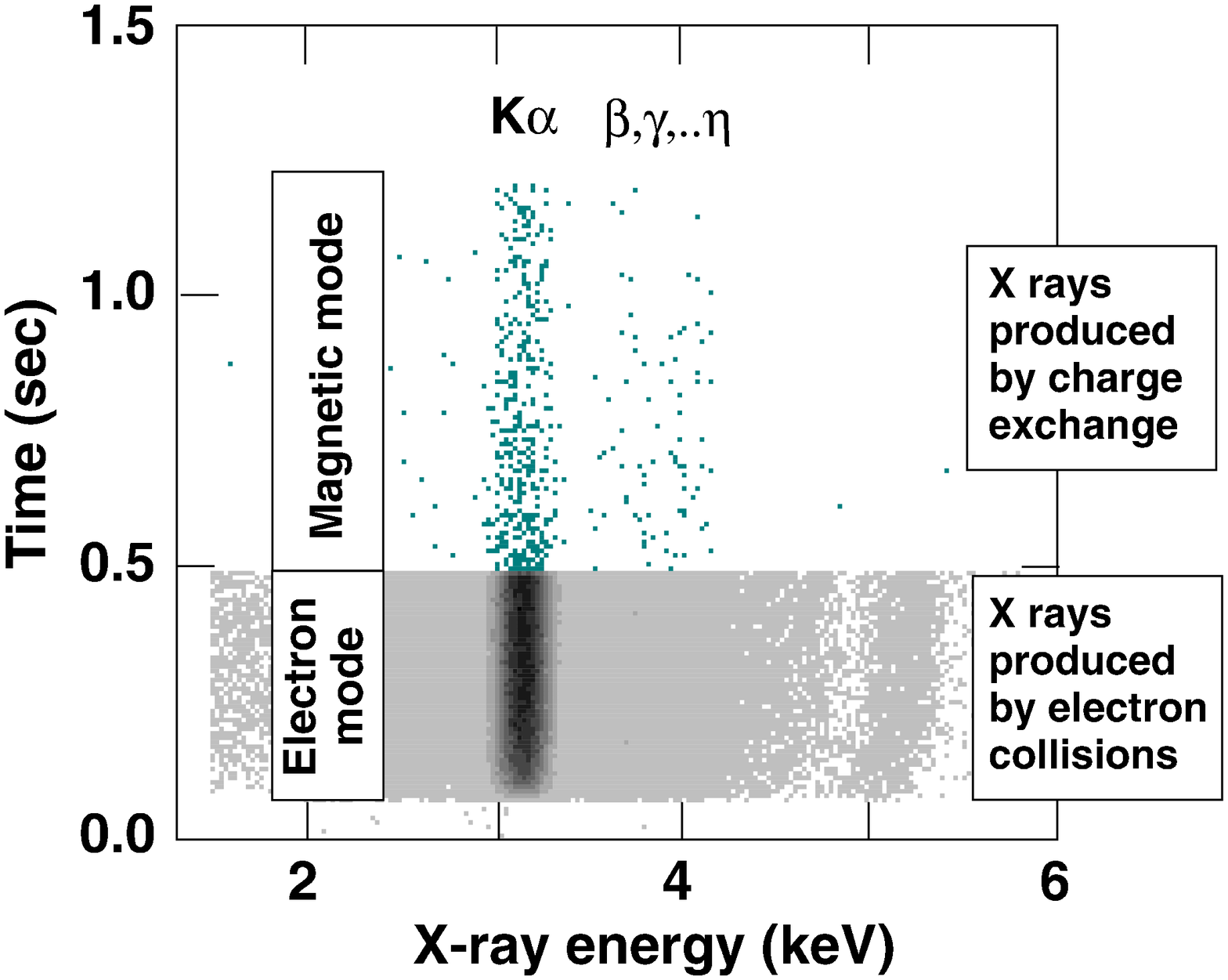}
\topcaption{
Time-resolved spectra of Ar obtained during the electron mode (beam on)
and magnetic mode (beam off) phases  
\protect\cite{cit:beiers2000prl85}.
The desired charge balance
(mostly He-like Ar$^{16+}$ with some H-like  Ar$^{17+}$) is established
with the beam on.  CX occurs during both phases but can only be seen
when the beam is turned off,
revealing He-like Ar K-series emission (\ka, \kb, ...)
from CX between Ar$^{17+}$ and neutral Ar.
He-like Ar$^{16+}$ ions also undergo CX, but the resulting Li-like emission
occurs below 1 keV.
\label{fig:magevent}
}
\end{figure}

In both cases the number of emitting ions, $N_{i}$, is roughly
the same; when the electron beam is on the ions are concentrated
around the electron beam, which defines the emitting volume,
and when the beam is off the ions expand into a larger cloud, but the
composition and total number of ions remains the same 
\cite{cit:beiersMag1994,cit:beiersMag1995}
and all of them
can undergo CX with the neutral gas.
There are large differences, however, 
in $n_{e}$ and $n_{n}$ (roughly $10^{12}$ versus $10^{6}$ cm$^{-3}$),
$\sigma_{DE}$ and $\sigma_{CX}$ ($10^{-20}$ versus $10^{-14}$ cm$^{2}$),
and $v_{e}$ and $v_{i}$ ($10^{10}$ versus $10^{6}$ cm/s),
yielding a net DE rate typically 
several thousand times higher than the rate for CX.
When conducting CX experiments,
the desired
charge state balance is established during the electron-beam-on phase
(electron mode) and then the beam is turned off 
(magnetic mode) to collect the CX spectrum.
Example time-resolved spectra are shown in Figure~\ref{fig:magevent}
\cite{cit:beiers2000prl85}.


When the electron beam is on, electron-ion collisions are essentially
unidirectional and the ions' emission is generally polarized.
During the magnetic-mode phase, however, ions and neutral gas atoms
or molecules collide in random directions and  the emission is
unpolarized.  Although the effects of polarization are routinely
taken into account during data analysis, their absence in CX
measurements makes analysis easier.  The absence of systematic
errors introduced by polarization corrections is also beneficial
because CX rates are low and statistical uncertainties are
relatively large compared to those in DE measurements.

Another difference from DE experiments is that dispersive spectrometers
are of little use in collecting CX spectra.  The reason is that CX
emission arises from throughout the relatively large ion cloud
in the trap, several hundred $\mu$m in diameter.  
Such a wide source can not be used with
grating and Bragg-crystal spectrometers, which 
rely on the narrowness of the electron beam (60-$\mu$m) in lieu of a slit.
The CX emission region is so wide that the effective spectrometer
resolution is severely degraded, and the use of a slit would block
too large a fraction of emission from entering the spectrometer.
Non-dispersive detectors, such as Ge or Si(Li) solid-state detectors
with resolutions of order 100 eV FWHM at 1 keV, are thus essential
for CX measurements.

The recent development of X-ray microcalorimeters 
[e.g., \citen{cit:silver2002,cit:mccammon2002,cit:kelley2007}]
for use in space missions and the lab has proven to be
a great help to CX studies, as their superior energy resolution
(a few to several eV) allows individual emission lines to be resolved,
which is particularly important in disentangling the closely spaced
high-$n$ emission lines that are a prominent feature in many
CX spectra (see Section~\ref{sec:results-hlike}).
The Lawrence Livermore EBIT group is fortunate to have a
collaboration with the Goddard Space Flight Center's X-Ray Spectrometer (XRS)
group, who have provided a 
$6\times6$-element array microcalorimeter with 32 active pixels
for dedicated
use on the LLNL EBITs \cite{cit:porter2004}.  
The XRS is described elsewhere in this volume (F.S.\ Porter, this issue).

\section{Results from Experiment and Theory}
\label{sec:results}


\subsection{$n$ and $l$ Distributions}
\label{sec:results-nldistrib}

\begin{figure}[t]
\centering
  \begin{minipage}[t]{0.47\textwidth}
    \centering
\includegraphics[angle=0,width=0.95\textwidth]{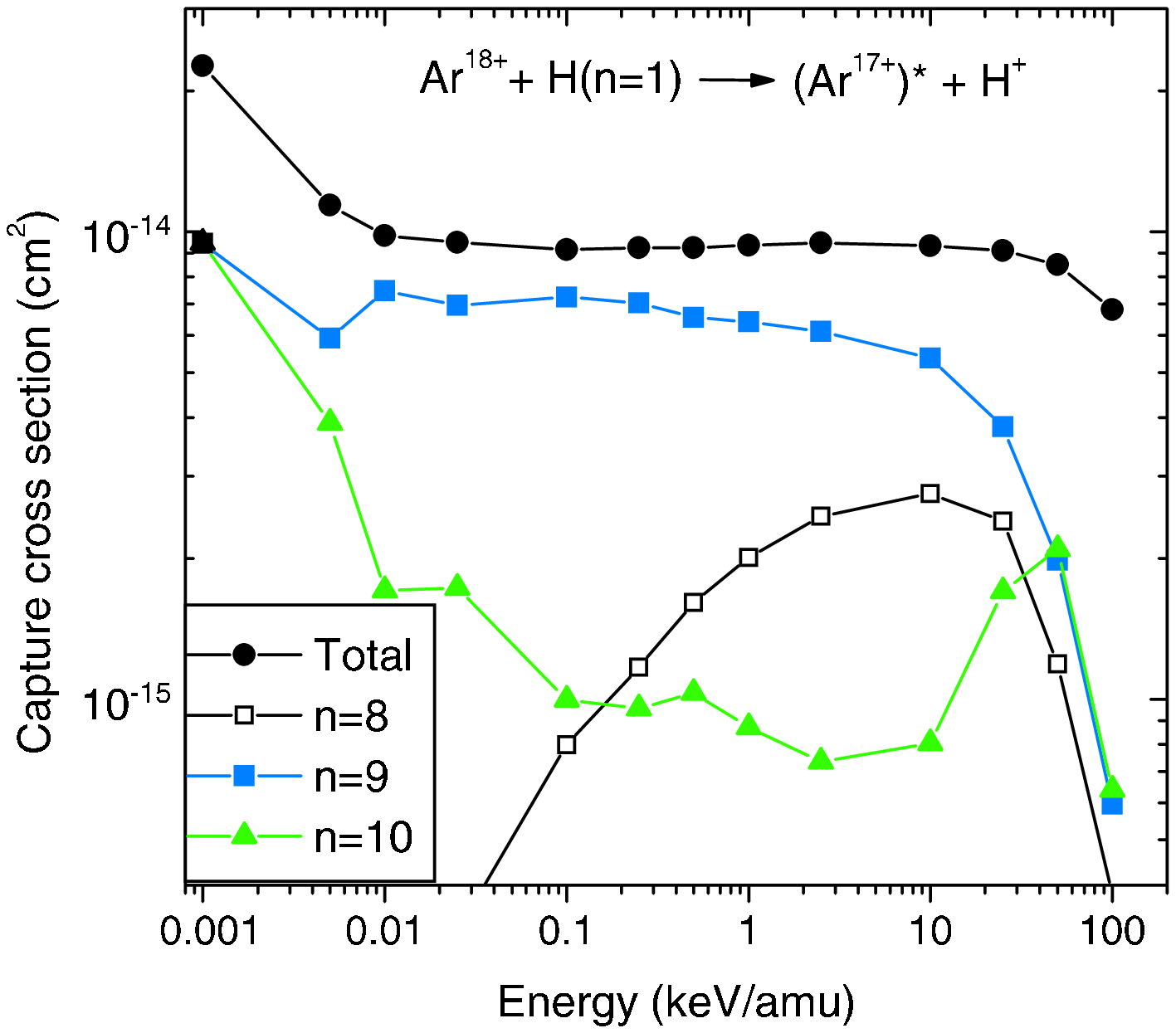}
\topcaption{
CTMC model predictions for Ar$^{18+}$ $+$ H, showing
cross sections for the most important $n$ levels.
Adapted from Perez et al.\ \protect\cite{cit:perez2001}.
\label{fig:theoNlevel}
}
\end{minipage}
  \hfill
%
\begin{minipage}[t]{0.47\textwidth}
\centering
\includegraphics[angle=0,width=0.95\textwidth]{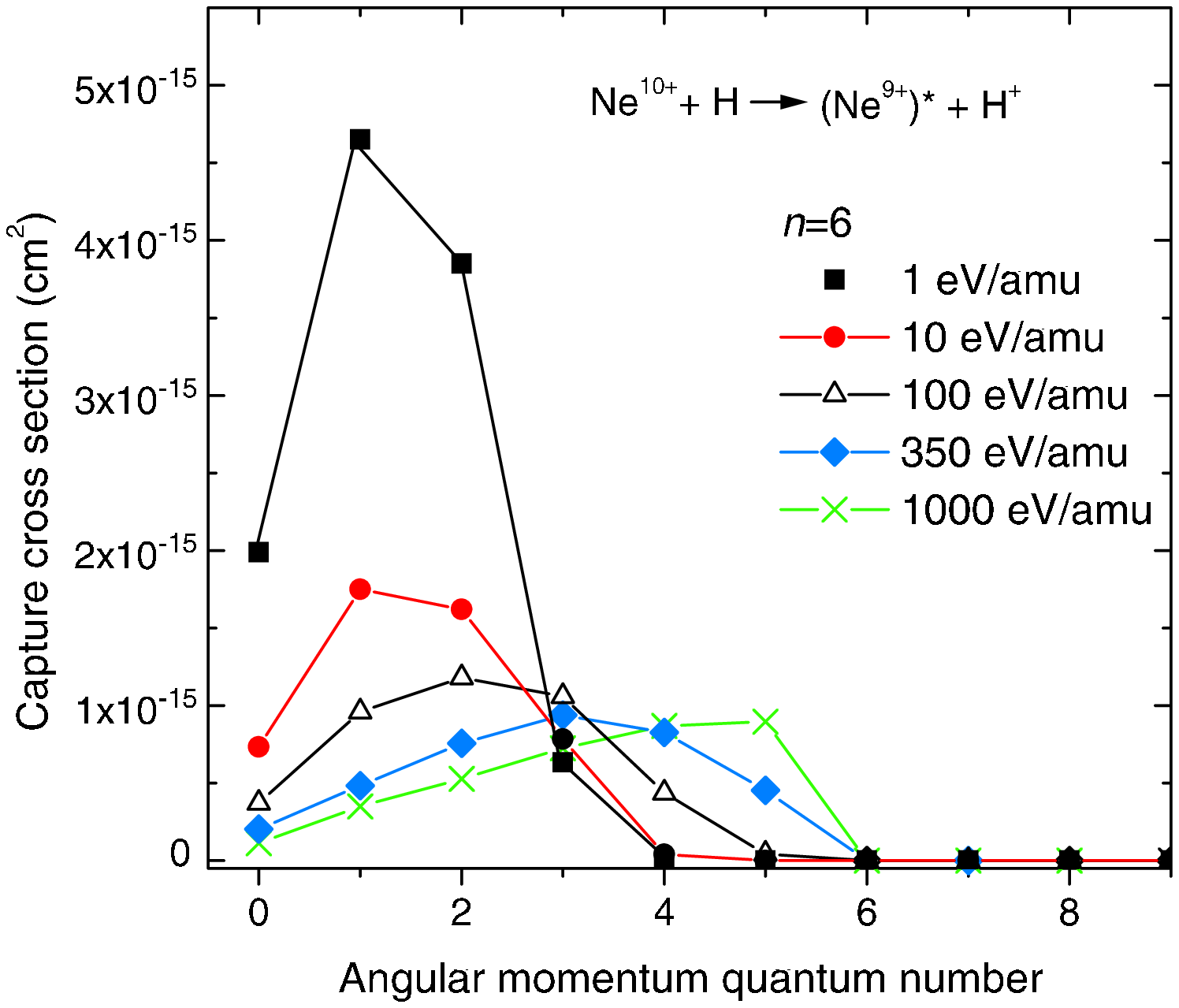}
\topcaption{
Model predictions for the $l$ distribution resulting
from CX of Ne$^{10+}$ and H.  Adapted from \protect\cite{cit:perez2001}.
\label{fig:theoLlevel}
}
\end{minipage}
\end{figure}

As seen in Figure~\ref{fig:kmjcross}, CX cross sections are roughly
constant over a wide range of energy.  
As discussed in Section~\ref{sec:basics-eqns},
the principal quantum number with the highest likelihood of
being populated by electron capture, $n_{max}$, also has little
energy dependence.  Theoretical results for the dominant levels 
populated by the CX of bare Ar$^{18+}$ and H are shown in 
Figure~\ref{fig:theoNlevel} \cite{cit:perez2001}.
As can be seen, the $n=9$ level dominates until energies
near $E_{crit}$ ($\sim$100 keV/amu).

In contrast to the behavior of the total cross section and $n_{max}$,
the angular momentum distribution has a strong dependence on energy,
as seen in Figure~\ref{fig:theoLlevel} 
for the $n=n_{max}=6$ level of H-like Ne following
CX between Ne$^{10+}$ and H \cite{cit:perez2001}.  
At energies below roughly 100 eV/amu,
the $l$-distribution peaks around $l=1$
while at energies above $\sim$1 keV/amu the $l$-distribution is
nearly statistical with $2l+1$ weightings out to the maximum $l$
value of $l_{max}=n-1$.

%


\subsection{H-like Spectra}
\label{sec:results-hlike}

\begin{figure}[t]
\centering
 \begin{minipage}[t]{0.47\textwidth}
    \centering
\includegraphics[angle=0,width=0.95\textwidth]{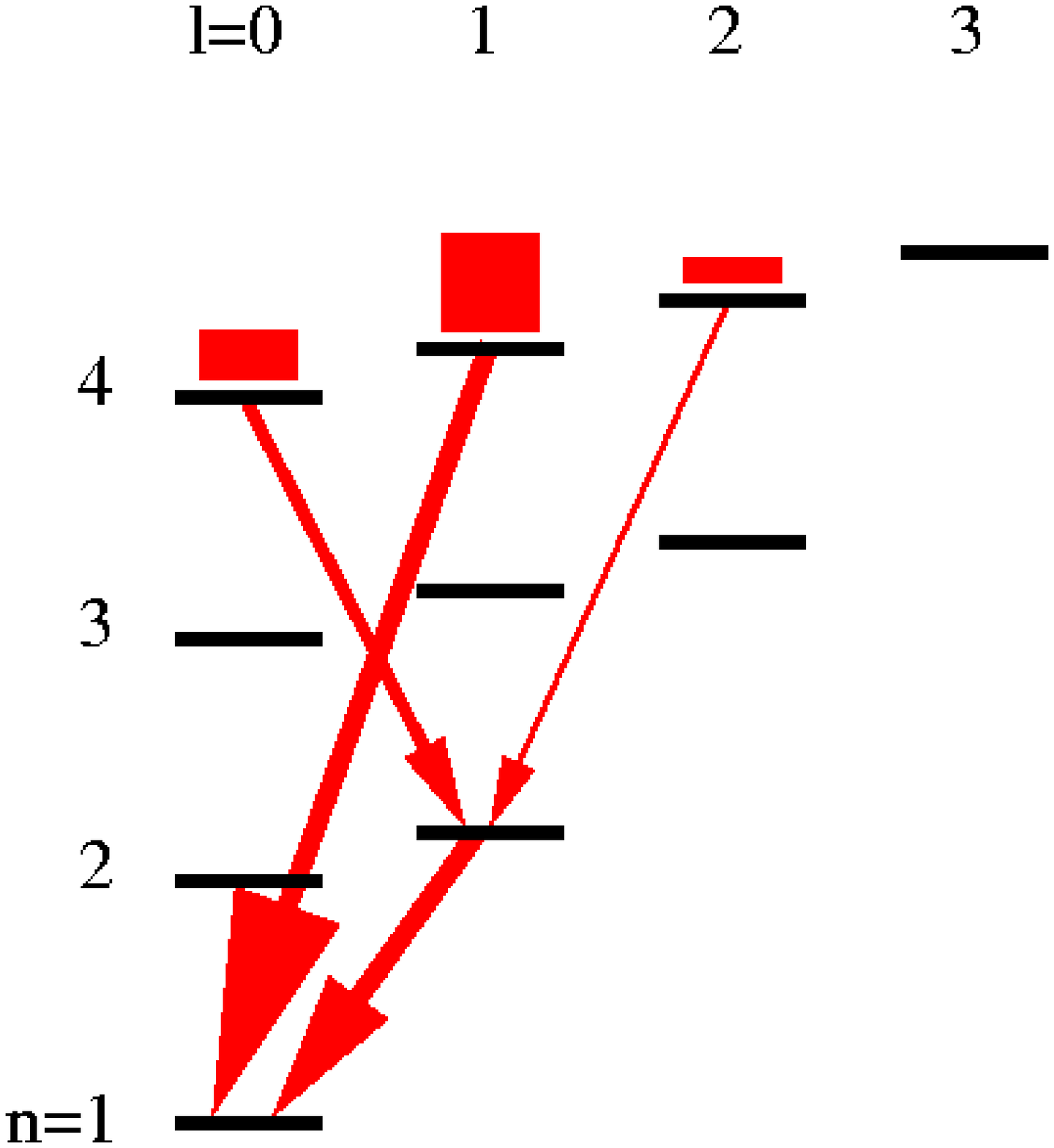}
\topcaption{
Schematic energy level diagram for a hydrogenic ion
illustrating radiative decay paths
following CX by a bare ion at low energies.  Immediately
following electron capture, the resulting H-like ion has
a large fraction of states with $l=1$ ($p$ states) that
can decay directly to ground, producing strongly
enhanced high-$n$ Lyman emission.
\label{fig:HlevelDecayLowE}
}
\end{minipage}
 \hfill
\begin{minipage}[t]{0.47\textwidth}
    \centering
    \includegraphics[angle=0,width=0.95\textwidth]{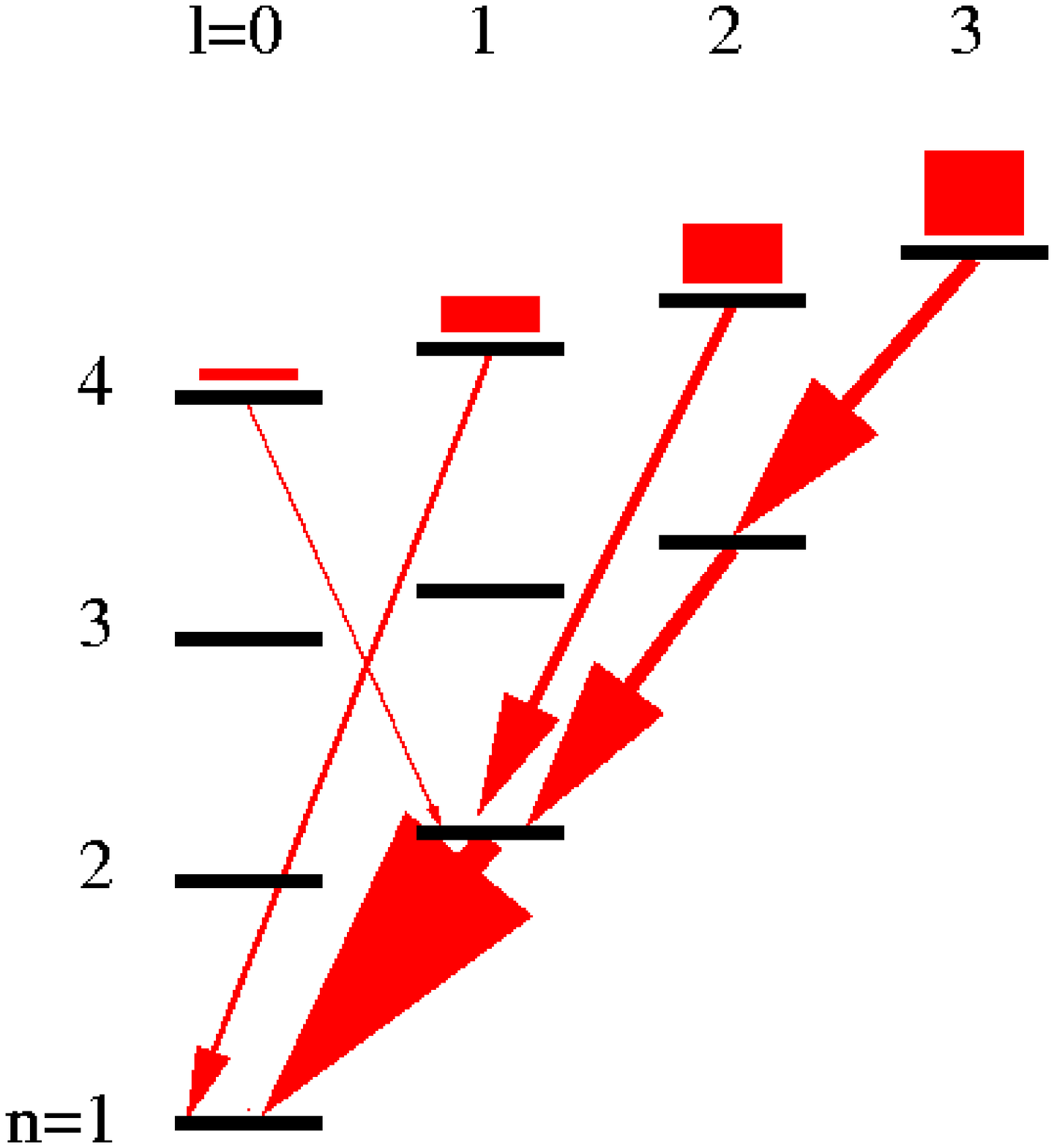}
\topcaption{
Radiative decay paths
following CX by a bare ion at high energies.  The initial
$l$ distribution is proportional to $2l+1$ and
most electrons decay to intermediate states with $l=l_{max}$ that then
cascade to ground, ultimately releasing a \lya\ photon.
\label{fig:HlevelDecayHighE}
}
  \end{minipage}
\end{figure}

Even more than the $n$-distribution, the $l$-distribution following
electron capture is
what determines the form of the emitted spectrum.
This is especially true for H-like spectra resulting from 
CX between a fully stripped ion and a neutral target.
At low energies, the captured high-$n$ electron is likely to 
be in a $p$ state, from which it can decay directly to
the $1s$ ground state yielding a Lyman photon
(see Figure~\ref{fig:HlevelDecayLowE}; the level for electron
capture is $n=4$ in this example).

At higher energies, higher angular momentum states are preferentially
populated, but because of the $\Delta l = \pm 1$ selection rule
for electric dipole (E1) radiation,
those states can not decay directly to ground.
Instead, the high-$n$ electron tends to decay to the lowest
$n$ level permitted by selection rules 
(see Figure~\ref{fig:HlevelDecayHighE}).  
In most cases, especially if the initial $n$ level is large,
the first decay will be to an $l$ state where
$l = l_{max} = n-1$, i.e., the maximum angular momentum state
for the given $n$ level.  From there the electron will
decay in $\Delta n = \Delta l = -1$ steps in a
`yrast cascade,'\footnote{
Yrast is the superlative of the Swedish word, yr, and means
`whirlingest,' or `dizziest.'  In atomic and nuclear physics
it refers to the highest angular momentum state for a given
principle quantum number.  An example of a yrast cascade
is $5g \rightarrow 4f \rightarrow 3d \rightarrow 2p \rightarrow 1s$.
}
ultimately releasing a \lya\ photon.
Only the small fraction of electrons initially captured
into $p$ states (along with some electrons that decay to
intermediate-$n$ $p$ states) decays directly to ground, so
the fraction of high-$n$ Lyman lines is very small.

\begin{figure}
  \centering
%
  \begin{minipage}[t]{0.48\textwidth}
    \centering
\includegraphics[angle=0,width=0.95\textwidth]{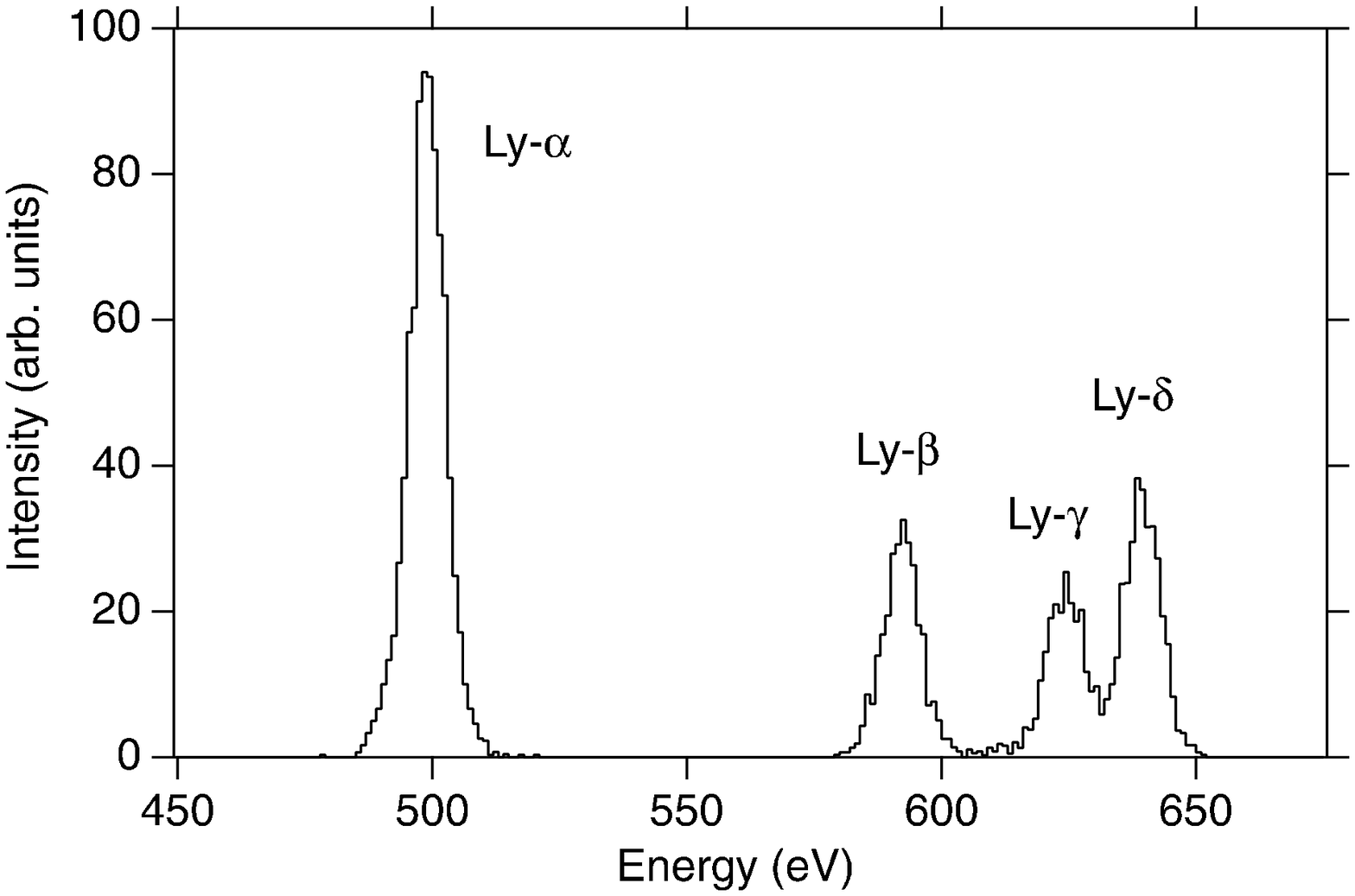}
\topcaption{
H-like spectrum of N$^{7+}$ $+$ CO$_{2}$ (adapted from
\protect\cite{cit:brownN2006}),
obtained using the XRS.
\label{fig:Nhlike}
}
\end{minipage}
  \hfill
%
  \begin{minipage}[t]{0.48\textwidth}
    \centering
\includegraphics[angle=0,width=0.95\textwidth]{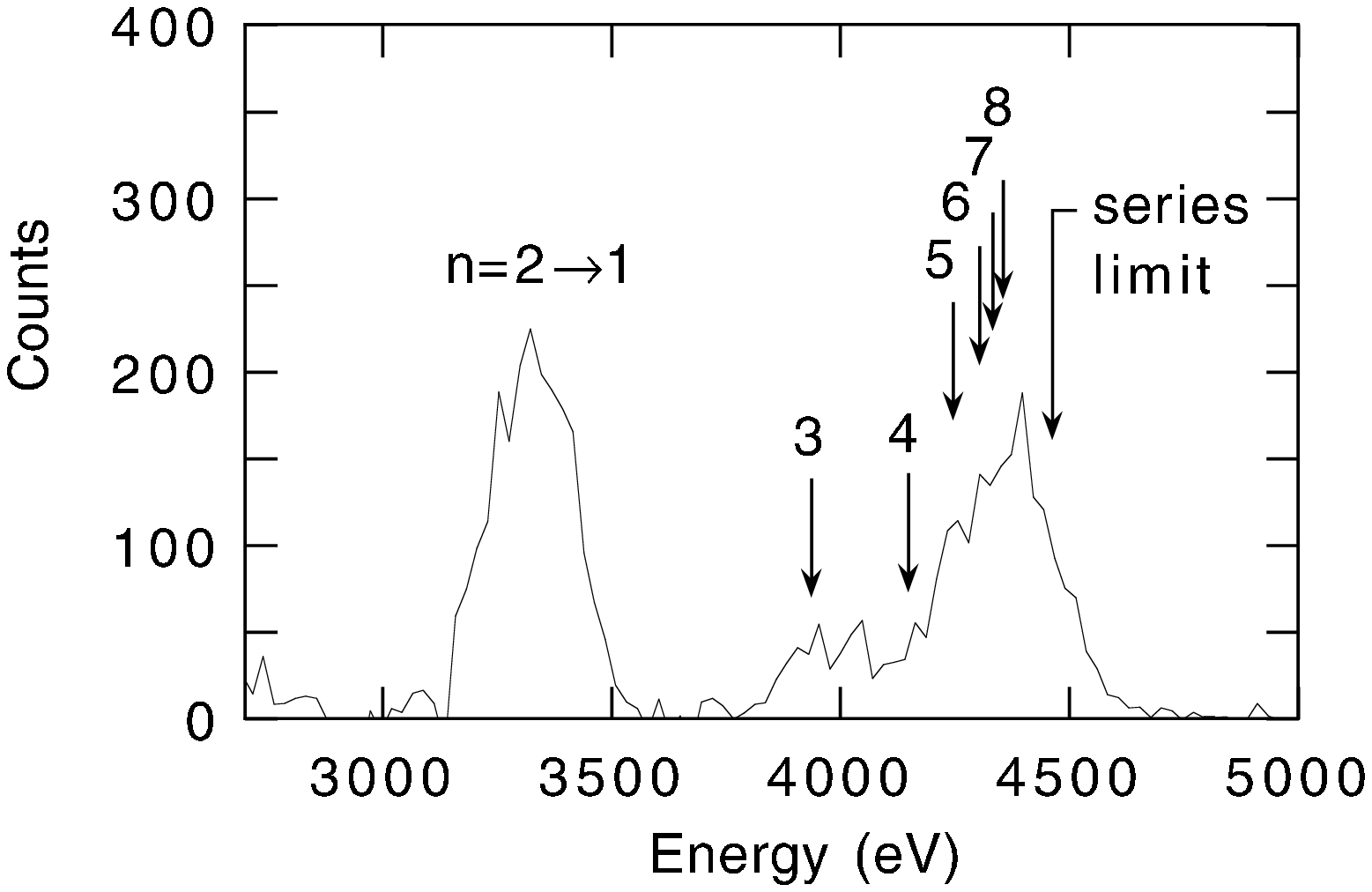}
\topcaption{
H-like spectrum of Ar$^{18+}$ $+$ Ar.  Adapted from
\protect\cite{cit:beiersArAu2000}.
\label{fig:ArHlike}
}
\end{minipage}
\end{figure}

\begin{figure}[]
  \centering
%
  \begin{minipage}[t]{0.48\textwidth}
    \centering
\includegraphics[angle=0,width=0.95\textwidth]{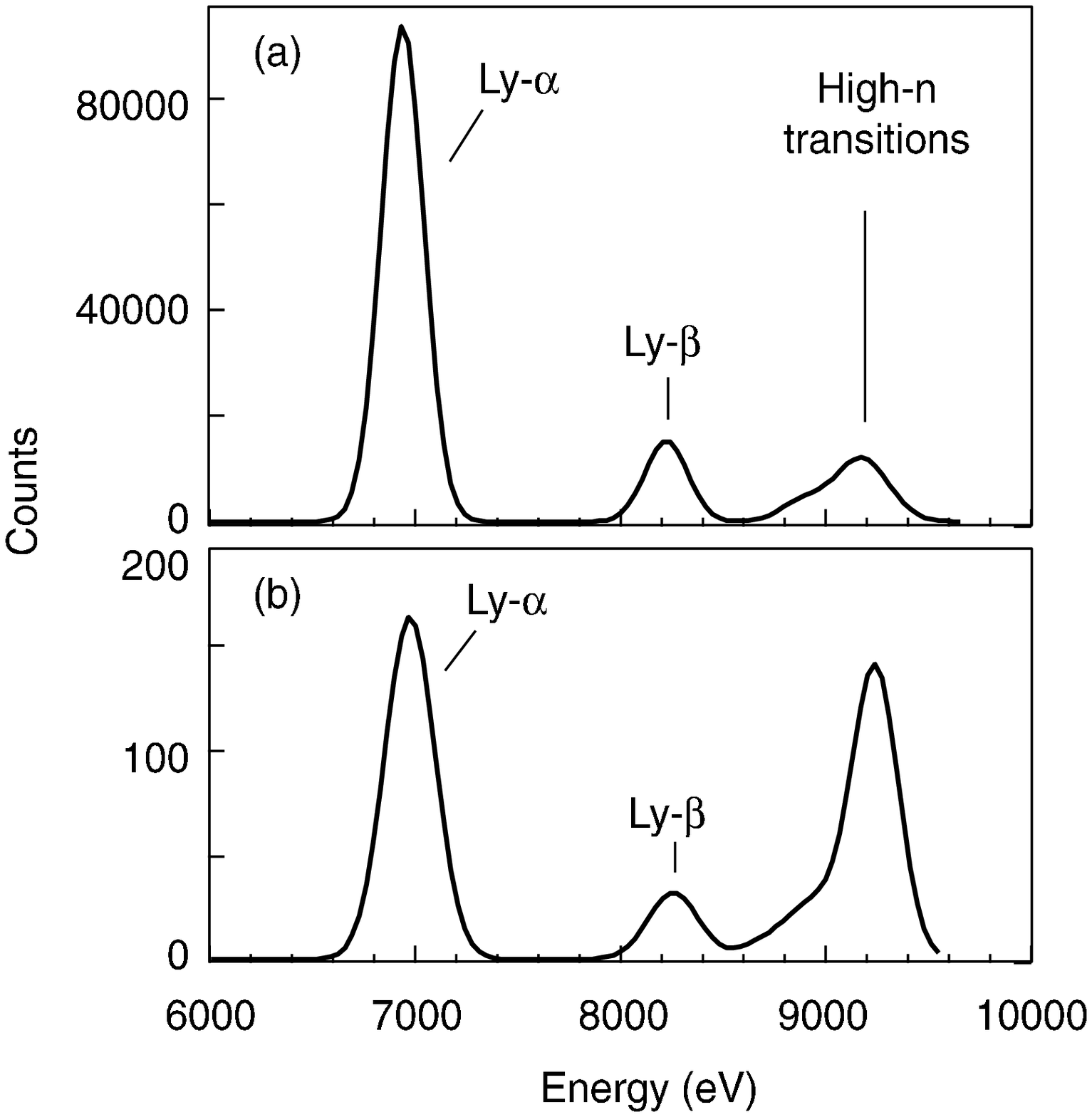}
\topcaption{
EBIT H-like Fe spectra from direct excitation (top panel)
and CX of Fe$^{26+}$ $+$ N$_{2}$ (bottom;
adapted from \protect\cite{cit:wargelin2005}).
High-$n$ Lyman emission is strongly enhanced in CX spectra
at low collision energies.
\label{fig:FeHlike}
}
\end{minipage}
\hfill
%
\begin{minipage}[t]{0.48\textwidth}
  \centering
  \includegraphics[angle=0,width=0.95\textwidth]{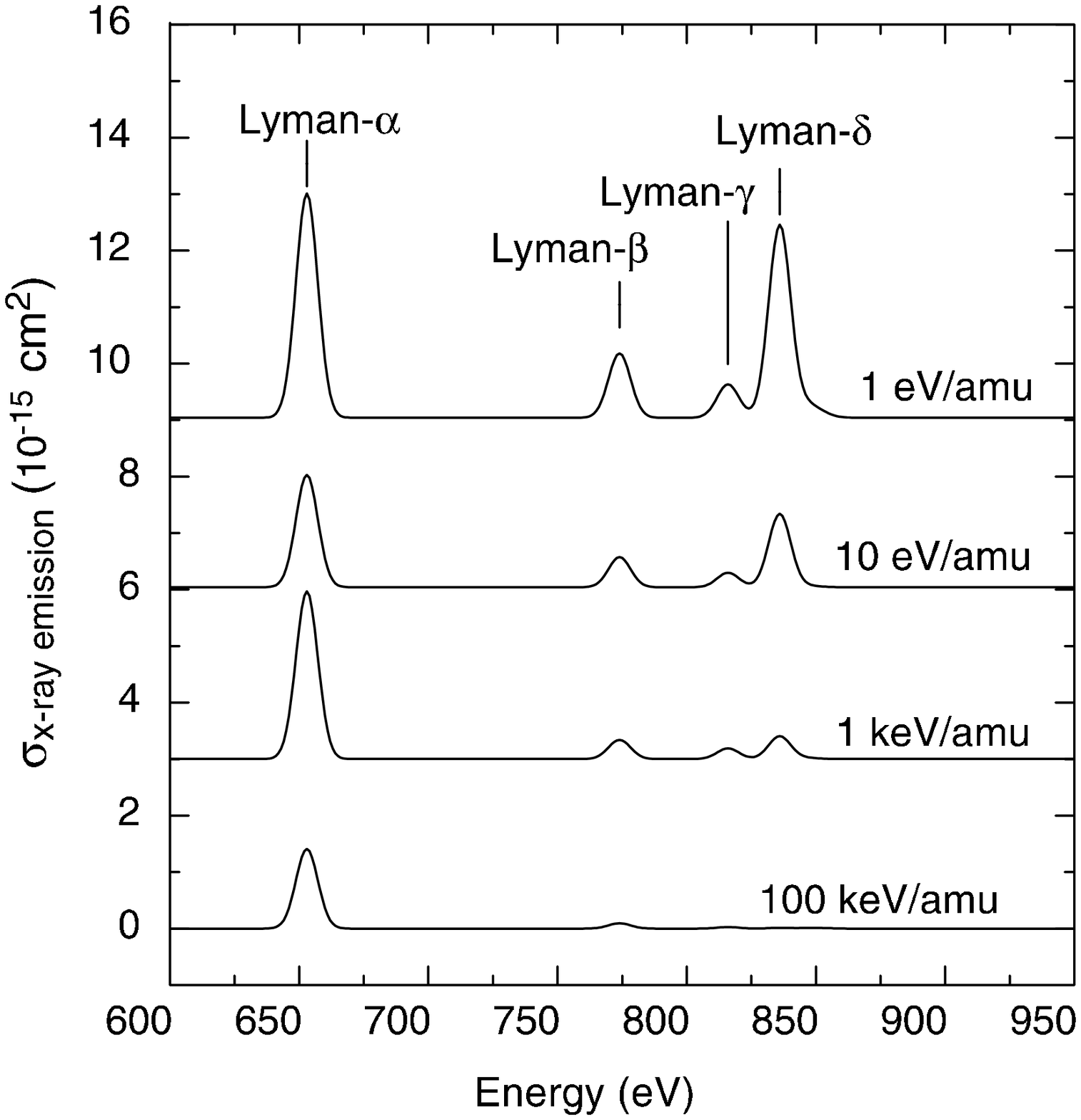}
\topcaption{
CTMC model spectra from CX of O$^{8+}$ and H at energies from
1 eV/amu to 100 keV/amu (adapted from Otranto et al.\ 
\protect\cite{cit:otranto2006}).  
High-$n$ emission is strongly enhanced
at low energies but nearly vanishes as the collision energy
approaches $E_{crit}$.
\label{fig:spectraFofE}
}
\end{minipage}
\end{figure}

Examples of EBIT spectra are shown for 
H-like N (N$^{7+}$ $+$ CO$_{2}$; \cite{cit:brownN2006})
and Ar (Ar$^{18+}$ $+$ Ar; \cite{cit:beiersArAu2000})
in Figures~\ref{fig:Nhlike} and \ref{fig:ArHlike}.
Note the large intensities of
the high-$n$ Lyman lines relative to \lya, especially
in the H-like Fe spectrum \cite{cit:wargelin2005}
of Figure~\ref{fig:FeHlike} 
which shows a DE spectrum for comparison.
This strong enhancement of high-$n$ Lyman emission is
a key feature of low-collision-energy CX spectra.
H-like CX spectra have also been obtained using EBITs for
Xe$^{54+}$ $+$ Xe \cite{cit:beiersMag1996,cit:perez2001},
Ne$^{10+}$ $+$ Ne \cite{cit:beiersVeloc2001,cit:beiersNe2005},
and O$^{8+}$ $+$ various molecules \cite{cit:beiersSci2003};
the last two measurements used the XRS microcalorimeter.

The energy dependence of H-like CX spectra is illustrated
in Figure~\ref{fig:spectraFofE} with theoretical results for O$^{8+} + $H
collisions at energies between 1 eV/amu and 100 keV/amu
\cite{cit:otranto2006}.
As can be seen, the fraction of high-$n$ emission decreases
at higher energies, and high-$n$ emission is negligible
above a few keV/amu (with a faster fall-off
for higher-$Z$ elements \cite{cit:perez2001}).
The hardness ratio 
$\mathcal{H} = (\mathrm{Ly}\beta + \mathrm{Ly}\gamma + \mathrm{Ly}\delta 
	+ \mathrm{Ly}\epsilon + ...) / \mathrm{Ly}\alpha$ 
can therefore be used as a collision energy diagnostic.

\subsection{He-like Spectra}
\label{sec:results-helike}

He-like spectra resulting from the CX of H-like ions
also have unique features
that distinguish them from DE spectra.
These spectra are somewhat more complicated than H-like spectra because
the spins of the two electrons
in He-like ions couple to yield either a singlet ($S=0$) or
triplet ($S=1$) state.
A 1:3 probability ratio is usually assumed
based on the statistical weight of
the levels, but statistical weights have been shown to incorrect
for some circumstances both experimentally and in 
detailed theoretical calculations \cite{stancil97,bliek98}
for Be-like ions, which also have $S=0$ and $S=1$ states.

\begin{figure}
  \centering
  \begin{minipage}[t]{0.47\textwidth}
    \centering
    \includegraphics[angle=0,width=0.95\textwidth]
        {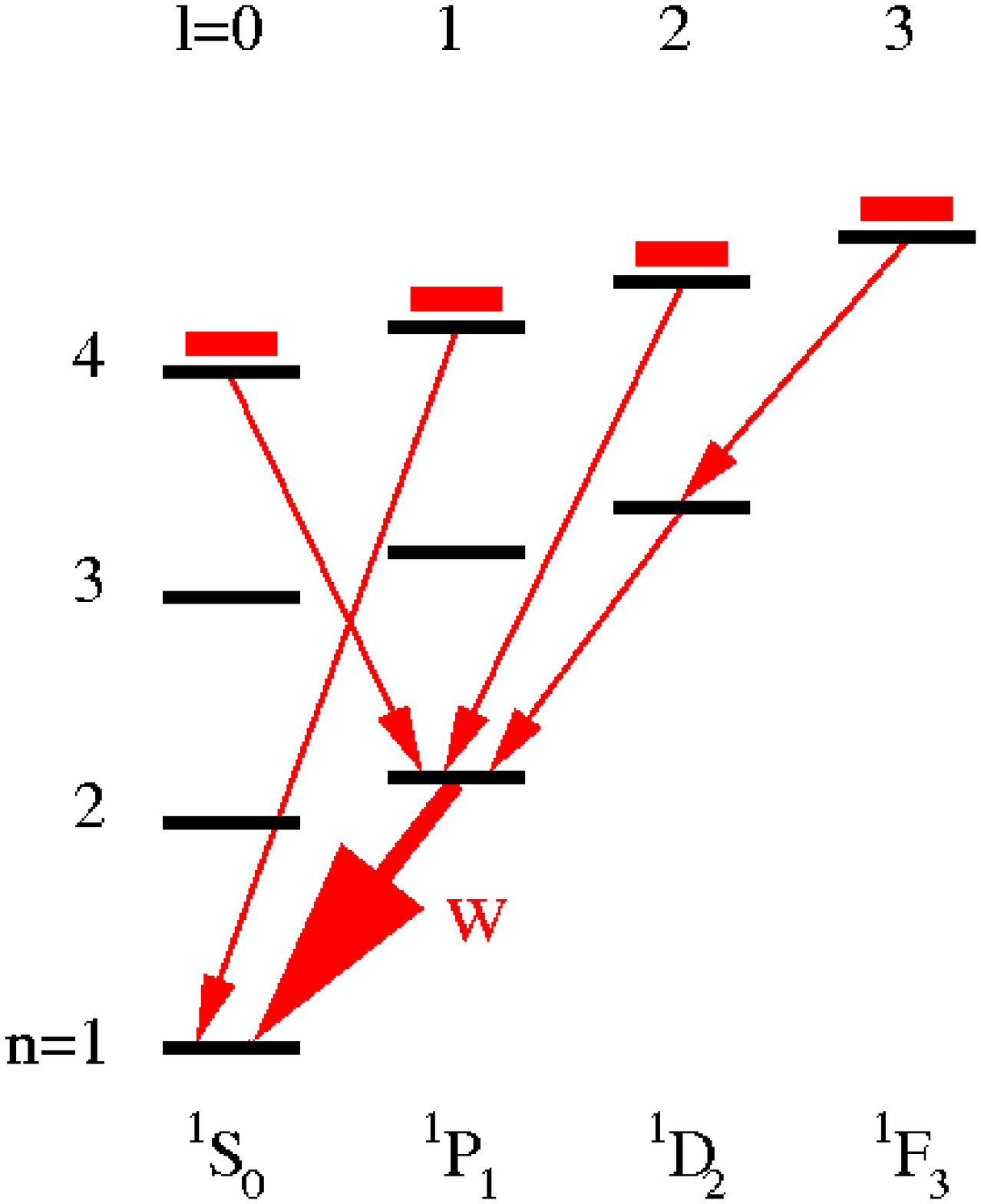}
  \topcaption{Energy level diagrams for He-like singlet ($S=0$)
	states.  
	He-like singlet-state CX spectra are similar to H-like spectra,
	with strong $n=2\rightarrow 1$ emission (line $w$) and significant
	high-$n$ emission at low collision energies.
	\label{fig:HeLevelSinglet}
	}
  \end{minipage}
  \hfill
  \begin{minipage}[t]{0.47\textwidth}
    \centering
    \includegraphics[angle=0,width=0.95\textwidth]
        {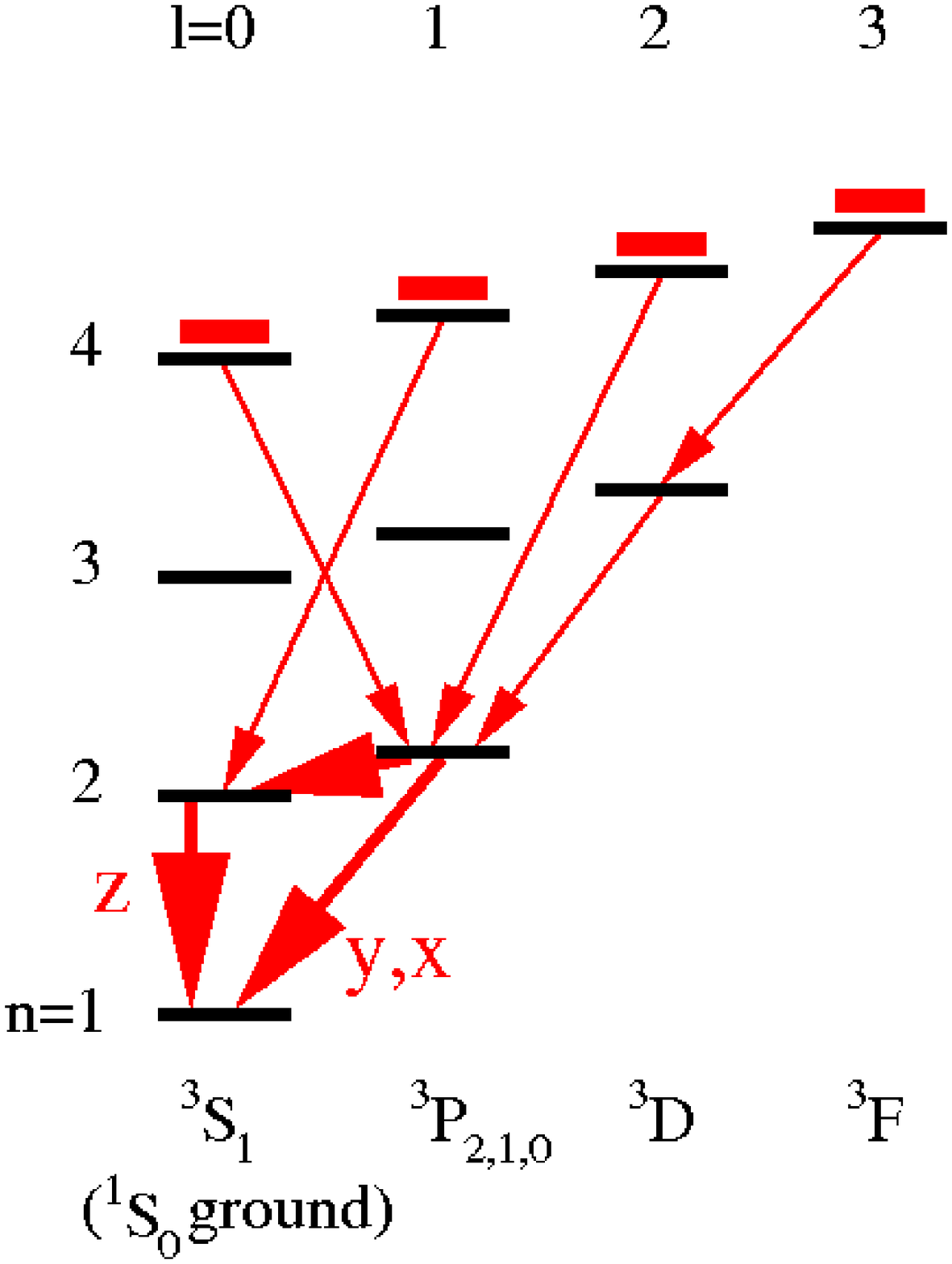}
  \topcaption{Energy level diagrams for He-like
	triplet ($S=1$) states (and the singlet $^{1}S_{0}$ ground state).
	High-$n$ states cannot decay
	to ground because of the $\Delta S=0$ selection rule,
	and instead decay to $n=2$ states from which they
	decay via forbidden or semi-permitted transitions 
	(the $z$, $y$, and $x$ lines),
	yielding \ka\ emission.
	\label{fig:HeLevelTriplet}
	}
  \end{minipage}
\end{figure}

Figures~\ref{fig:HeLevelSinglet} and \ref{fig:HeLevelTriplet}
show schematic energy level diagrams for electron capture
into $n=4$ singlet and triplet states, respectively,
assuming a uniform distribution of angular momentum states
for illustrative purposes.
The decay of the singlet states has essentially the same
behavior as for H-like ions, yielding
a prominent $n= 2\rightarrow1$ line
(specifically the $w$ `resonance' 
$1s2p\,^{1}P_{1} \rightarrow 1s^{2}\, ^{1}S_{0}$ transition, 
one of the four
components of the \ka\ complex; see below)
along with
some high-$n$ K-series emission (especially at low collision energies, as
is the case for H-like CX emission).


The triplet decay scheme is different, however, because
the $\Delta S= 0$ selection rule forbids mixing,
or `intercombination,' of triplet and singlet states.
As a result, {\it none} of the initial triplet states
can decay directly to ground.
Instead, all triplet states eventually decay to one of
the $n=2$ triplet states: $1s2s\,^{3}S_{1}$, or
$1s2p\,^{3}P_{2}$, $^{3}P_{1}$, and $^{3}P_{0}$.
From there they decay to ground via forbidden or
semi-permitted transitions.

Starting with the highest $n=2$ energy level and proceeding to
lower energies,
the $^{3}P_{2}$ state is forbidden to decay
to ground by the
$\Delta J =0,\pm1$ (but no $J=0 \rightarrow 0$) selection rule.
It can decay via a $\Delta J =2$
magnetic quadrupole (M2) transition (yielding the $x$ line,
an `intercombination' transition so-called because it
involves a change in $S$)
but usually decays to
the $^{3}S_{1}$ state.
The $^{3}P_{1}$ state mixes slightly with the
$^{1}P_{1}$ state in $J-J$ coupling and can therefore decay to ground
via another intercombination transition to produce the $y$ line.
The $^{3}P_{0}$ state is
strictly forbidden to decay by single-photon emission\footnote{
	The only exception is in isotopes with a non-zero nuclear 
	magnetic moment, which mixes 
	$^{3}P_{0}$ with the $~^3P_1$ state via the hyperfine interaction
	and thus enables decay 
	to ground (e.g., see \cite{wong95}).}
because of
the no $J=0 \rightarrow 0$ rule and instead feeds the $^{3}S_{1}$ state,
which is the lowest lying of all the triplet states.
This
metastable state eventually decays to ground
via relativistic magnetic dipole (M1) decay, yielding the
`forbidden' $z$ line.

\begin{figure}
  \centering
  \begin{minipage}[t]{0.47\textwidth}
    \centering
    \includegraphics[width=1.00\textwidth]
        {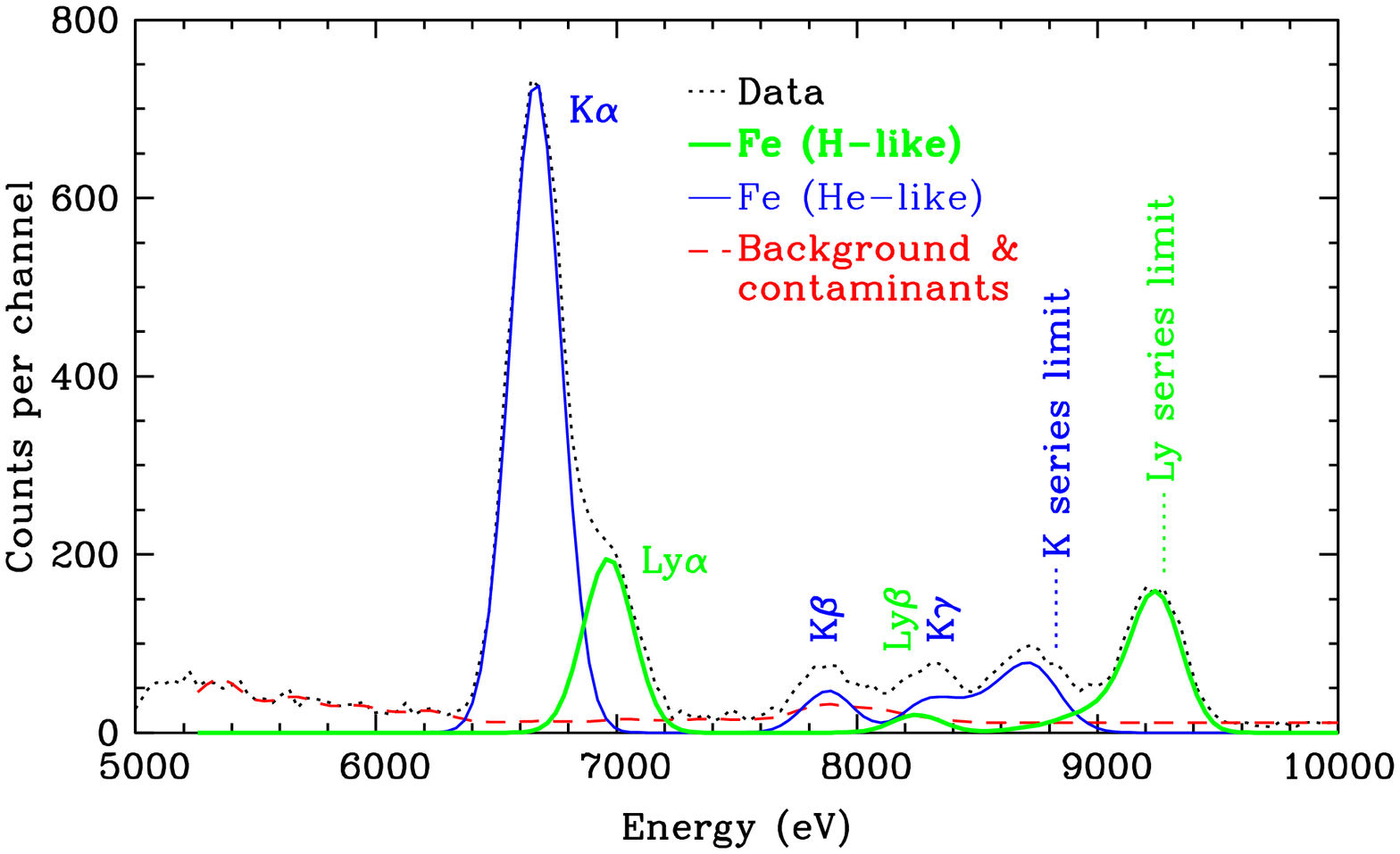}
  \topcaption{H-like and He-like spectra from CX of Fe ions with N$_{2}$
	(adapted from \protect\cite{cit:wargelin2005}).
	High-$n$ emission is more strongly enhanced in the H-like spectrum
	than in the He-like.
	\label{fig:FeCXwarg}
	}
  \end{minipage}
  \hfill
  \begin{minipage}[t]{0.47\textwidth}
    \centering
    \includegraphics[width=0.95\textwidth]
        {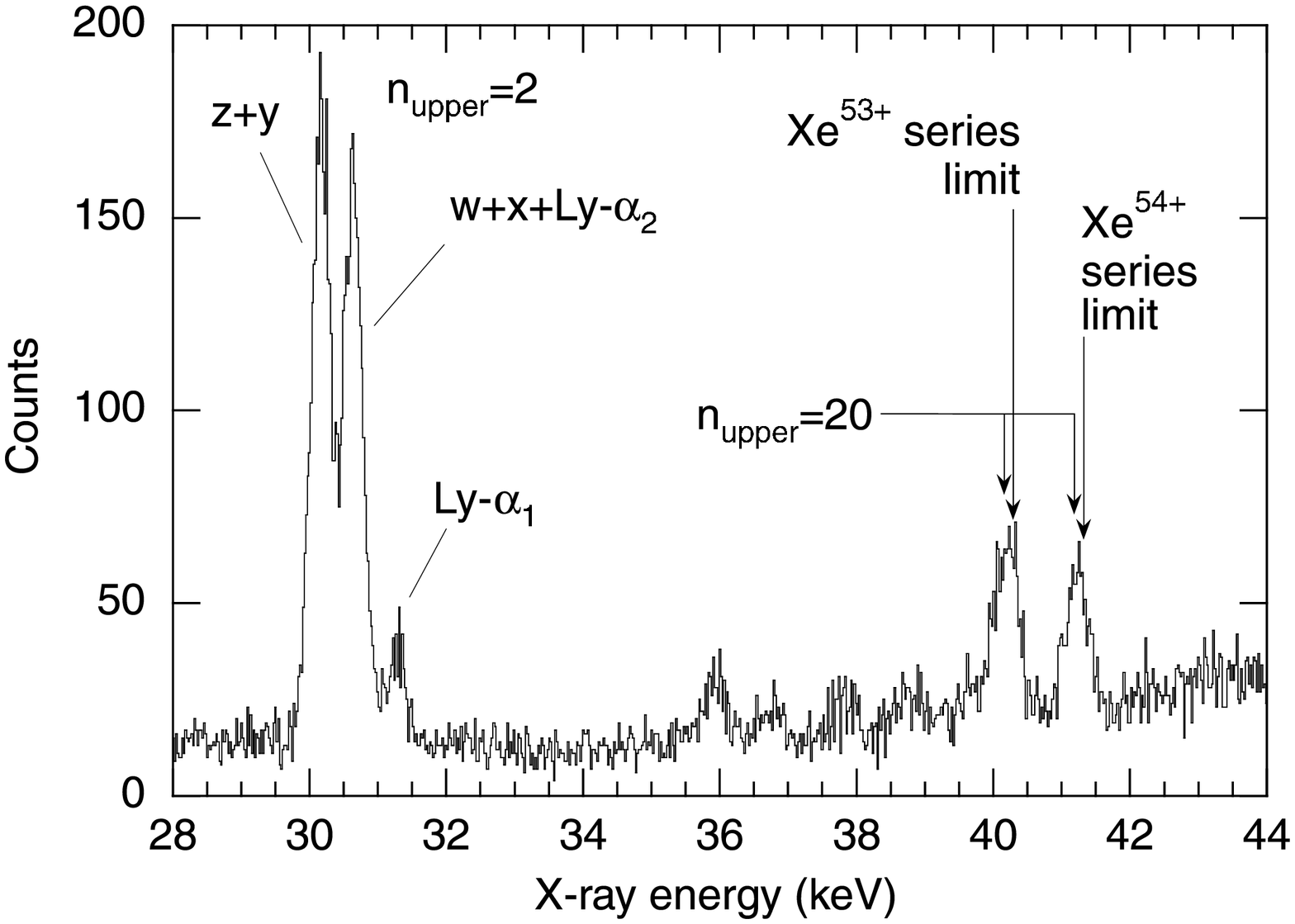}
  \topcaption{H-like and He-like spectra from CX of Xe ions with neutral Xe
	(adapted from \protect\cite{cit:perez2001}).
	\label{fig:XeCXperez}
	}
  \end{minipage}
\end{figure}

Most of the initial triplet states therefore end up
yielding the $z$ forbidden line or else the
$y$ intercombination line.  
Together with the weaker $x$ line and the $w$ resonance line, 
these make up the \ka\ complex, often but imprecisely referred to as the He-like
triplet (after $w$, the often blended $x$ and $y$, and $z$, 
and not to be confused with triplet, i.e., $S=1$, states).  
Because triplet-state
emission is usually several times as strong as singlet emission
and the triplet emission is completely dominated by \ka,
He-like CX spectra have weaker high-$n$ emission
than H-like spectra, although still stronger than in DE spectra.
He-like and H-like CX spectra of Fe and Xe are shown in
Figures~\ref{fig:FeCXwarg} and \ref{fig:XeCXperez};
He-like CX spectra have also been obtained on EBIT for
U$^{91+}$ $+$ Ne \cite{cit:schweikard1998},
Au$^{78+}$ $+$ Ne \cite{cit:beiersArAu2000},
Kr$^{35+}$ $+$ Kr \cite{cit:beiersKr1999},
Ar$^{17+}$ $+$ Ar \cite{cit:beiersArAu2000},
and Ne$^{9+}$ $+$ Ne \cite{cit:beiersArAu2000,cit:otranto2006} (the
latter with the XRS).



\begin{figure}
  \centering
    \includegraphics[width=0.65\textwidth]
        {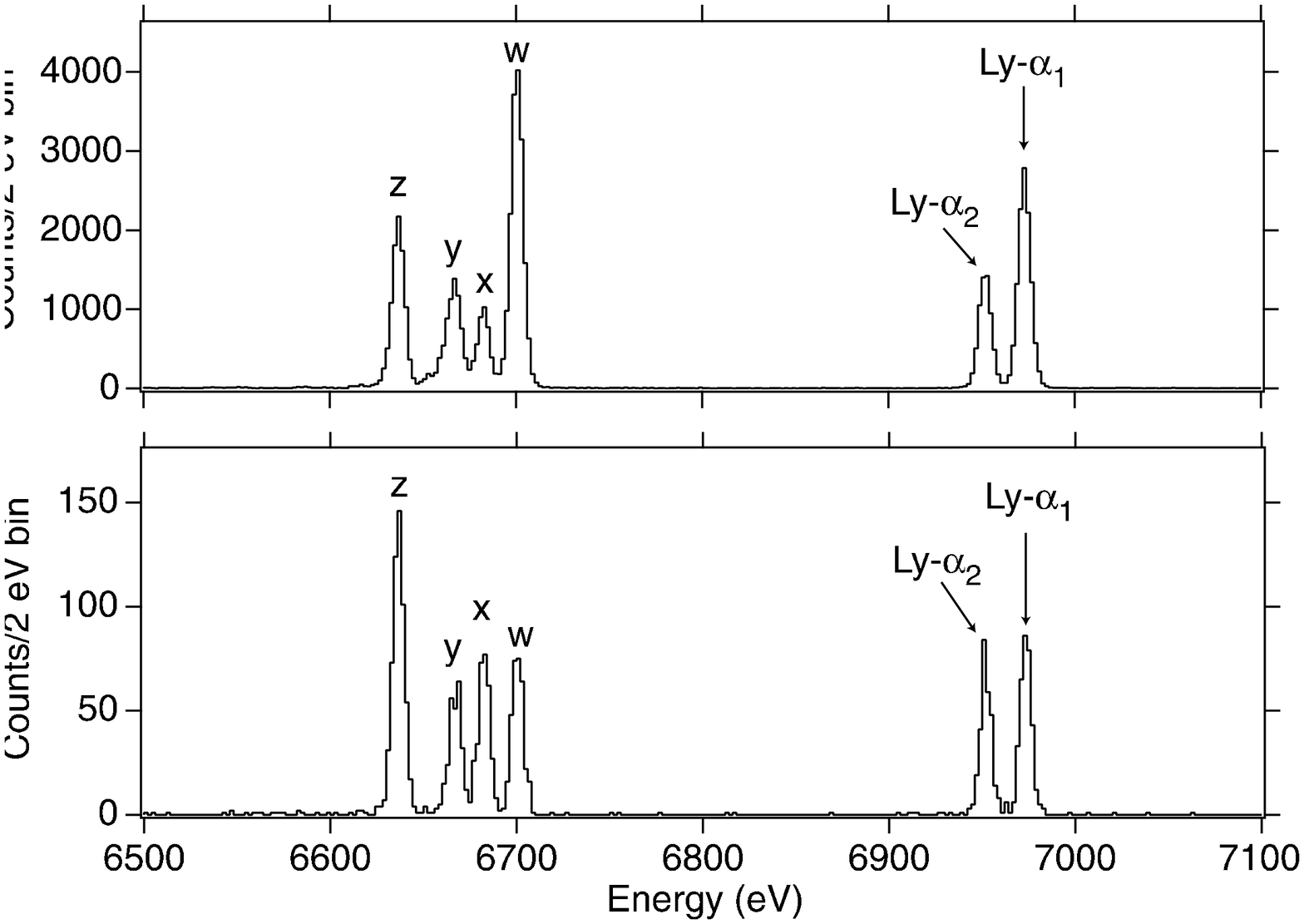}
  \topcaption{Experimental He-like and H-like Fe spectra from direct excitation 
	at 15 keV (top)
	and from CX with N$_{2}$ (bottom).  Both spectra were collected using
	the XRS microcalorimeter.  The $w$ line is relatively much weaker in
	the CX spectrum and the Ly$\alpha_1$/Ly$\alpha_2$ ratio is nearly
	unity instead of 2:1.
	\label{fig:HeFeCXvsDE}
	}
\end{figure}

Although there is relatively little enhancement of He-like high-$n$ lines,
there are still large differences between DE and CX spectra.
The much higher fraction of triplet states populated by CX
than by DE leads to large differences in the
relative intensities of the triplet and singlet lines,
as shown in Figure~\ref{fig:HeFeCXvsDE}.
Even if the $z$, $y$, $x$, and $w$ lines can not be resolved from each other,
the centroid energy of the \ka\ complex can often be determined
precisely enough to distinguish between a CX or DE origin.
As an example, based on the $\sim$20-eV difference in the centroid
energy of the He-like Fe \ka\ complex \cite{cit:wargelin2005},
Fujimoto et al.\ \cite{cit:fujimoto2007} concluded
from a \suzaku\ observation of diffuse emission in the Galactic Center
that cosmic-ray CX was not a significant contributor to
the observed signal (see Section~\ref{sec:astro-gc}).


\subsection{Multi-Electron Targets}
\label{sec:results-targets}

\begin{figure}
  \centering
  \begin{minipage}[t]{0.48\textwidth}
    \centering
    \includegraphics[width=0.95\textwidth]
        {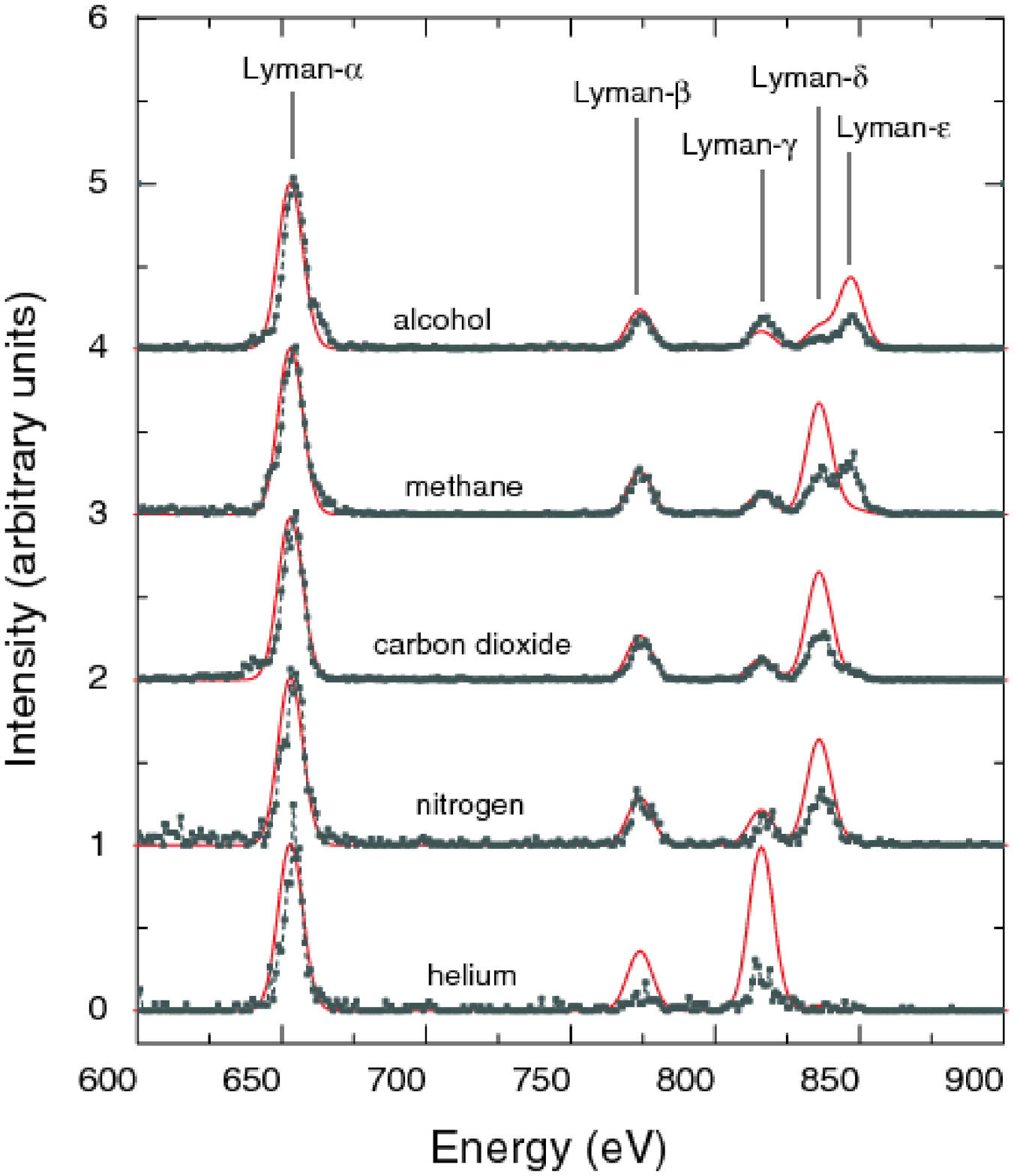}
  \topcaption{
	Experimental and CTMC model H-like O spectra from 
	CX of O$^{8+}$ with several neutral gases.
	The $n$ level of maximum likelihood for electron capture
	($n_{max}$) decreases as the target ionization
	potential increases approximately as expected,
	but the observed high-$n$ line intensities
	are generally smaller than predicted.
	Adapted from \protect\cite{cit:otranto2006}.
	\label{fig:otranto7}
	}
  \end{minipage}
  \hfill
  \begin{minipage}[t]{0.48\textwidth}
    \centering
    \includegraphics[width=0.95\textwidth]
        {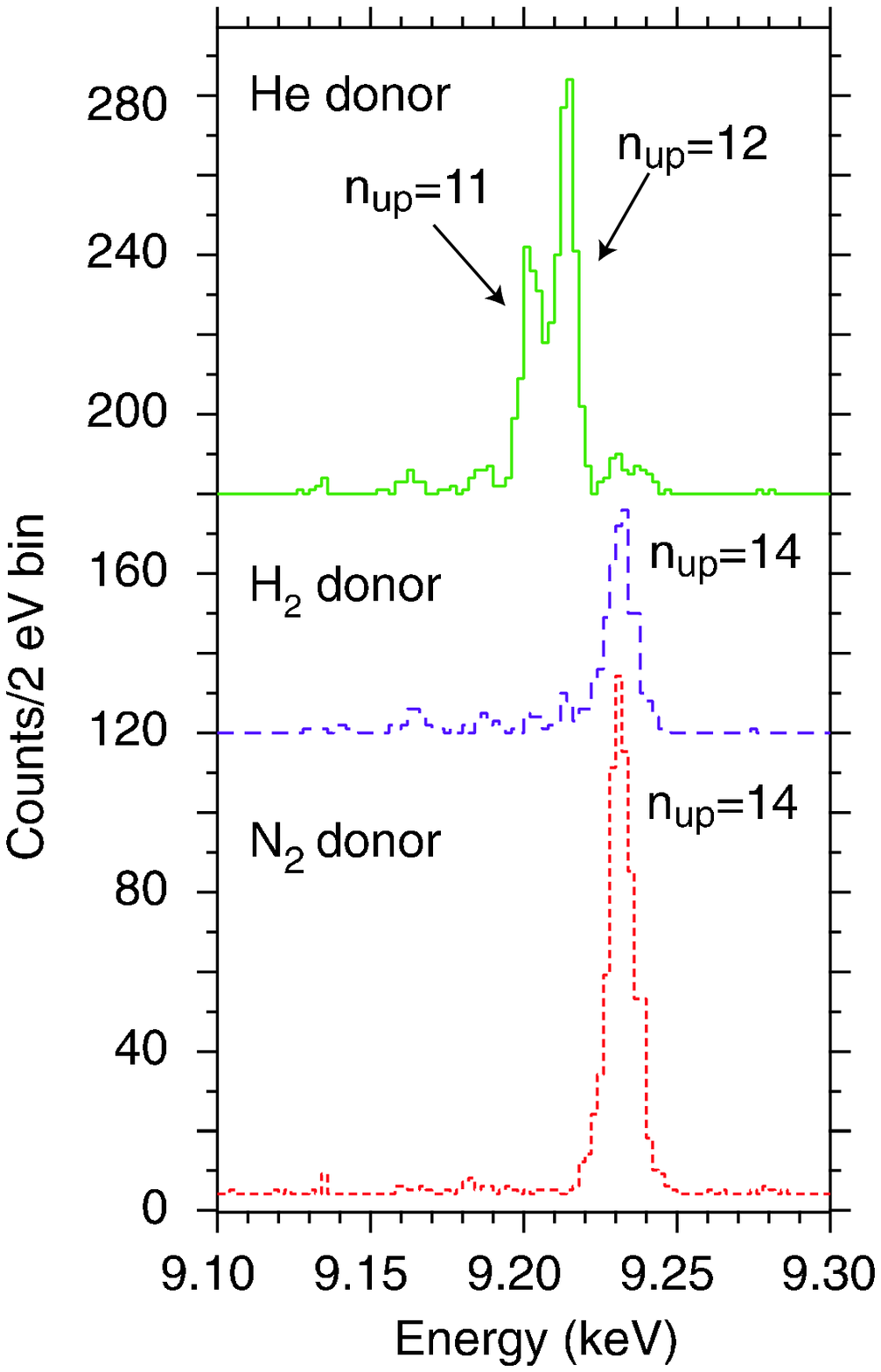}
  \topcaption{
	High-$n$ Lyman lines from measurements of
	CX of Fe$^{26+}$ with various neutral
	gases illustrating the effect of target ionization potential.
	Although the qualitative behavior is as expected,
	observed values of $n_{max}$ are larger than predicted by 
	extrapolations from CX with atomic H.
	\label{fig:FeHiN}
	}
  \end{minipage}
\end{figure}

Although the general features of CX spectra are largely independent
of the target species, the effects of differing ionization potentials
and multiple electrons 
can be quite pronounced, particularly
on high-$n$ emission lines.
Figure~\ref{fig:otranto7}
shows results from EBIT experiments and CTMC calculations
for CX of bare O$^{8+}$ with several targets 
of differing ionization potentials.
As expected from its $\sqrt{I_{H}/I_n}$ dependence
(see Eq.~\ref{eq:nmax}), $n_{max}$ decreases
as the ionization potential increases.
Similar behavior is seen in Figure~\ref{fig:FeHiN} 
for the high-$n$ lines of H-like Fe
for CX with N$_{2}$ (15.6 eV), H$_{2}$ (15.4 eV), and
He (24.6 eV).


Theoretical models are less successful, however, in
predicting line intensities.  Model results
using the Classical Trajectory Monte Carlo method
are shown in Figure~\ref{fig:otranto7}.  
Although the model usually does a good job
of predicting $n_{max}$ for the O spectra,
the intensities of high-$n$ lines relative to \lya,
especially for $n$ near $n_{max}$ is often poor.
For higher-$Z$ projectiles, discrepancies between
theory and experiment also appear in $n_{max}$.
In H-like Fe spectra
(Figure~\ref{fig:FeHiN}),
the predicted $n_{max}$ for
CX with N$_{2}$ and H$_{2}$ is 12 or 13, while the
observed value is 14 or 15.  For He the expected value
is 9 but the observed is 12.

\begin{figure}
    \centering
    \includegraphics[width=0.65\textwidth]
        {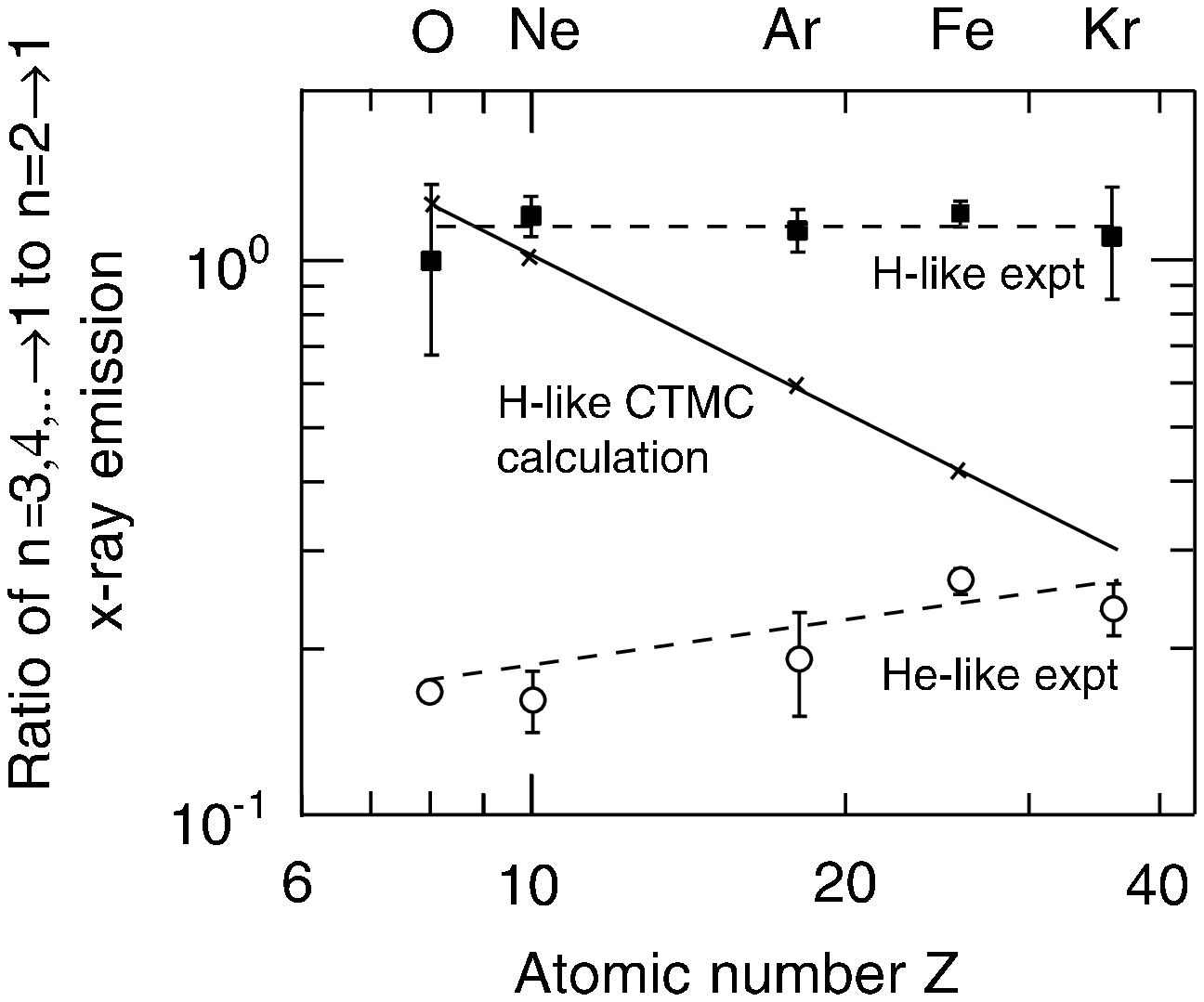}
  \topcaption{
	Hardness ratios for H-like and He-like CX emission as a function of $Z$,
for collision energies of $\sim$10 eV/amu.  Dashed lines through the H-like
measurements (solid points) and He-like measurements (open circles) are
drawn only to guide the eye.  Neutral gases used in the experiments are:
CO$_{2}$ (for O) \cite{cit:beiersSci2003}, 
Ne (for Ne) \cite{cit:beiersVeloc2001}, 
Ar (for Ar) \cite{cit:beiersArAu2000}, 
N$_2$ (for Fe) \cite{cit:wargelin2005}, 
and Kr (for Kr) \cite{cit:beiersArAu2000}.
Results from CTMC calculations ($\times$'s) for CX with atomic H
are extrapolated to Kr (solid line).
Adapted from \cite{cit:wargelin2005}.
	\label{fig:FeHR}
	}
\end{figure}

The overall intensity of high-$n$ emission with multi-electron targets
can also be a surprise.
Figure~\ref{fig:FeHR} plots the Lyman  series
hardness ratio ($\mathcal{H}$)
for CX using several combinations of bare projectiles and targets.
As can be seen, $\mathcal{H}$ is remarkably consistent.
Theoretical calculations for CX with the many-electron target gases used
in those experiments are not available,
but CTMC model calculations for single-electron CX with H
predict a steady decrease in $\mathcal{H}$ as $Z$ increases.
Multi-electron CX may be a significant process
in these collisions, but the resulting
autoionization of the multiply excited states
following MEC will shift the resulting high-$n$ Lyman emission
to lower values of $n$, thus making it even more difficult
to explain the higher than expected
values of $n_{max}$ that are observed in the Fe spectra. 
Whatever the contribution of MEC, it seems that
the presence of multiple target electrons somehow increases
the probability of capture into high-$n$ $p$ states that
can decay directly to ground.
The constancy of the hardness ratio across a wide range 
of H-like ions and many-electron targets suggests the existence of an 
underlying general process that perhaps could by approximated 
with relatively simple calculations.

Recent measurements at the Berlin EBIT \cite{allen07} have confirmed
the Livermore observation of the large hardness ratio observed for
Ar$^{17+}$ after interacting with neutral argon gas. The authors
also extracted Ar$^{18+}$ ions and, as in
crossed-beam experiments, had them interact with an argon gas target.
Surprisingly, the observed hardness ratio was much smaller and in
line with the CTMC calculation shown in Fig.\ 25.  This result
suggests that conditions in the crossed-beam
differ in some significant way from 
those in the EBIT measurement.
One difference might be the effect of fields on level-specific
CX cross sections,
however, lowering the field strength (3 T during normal operation for
both the Berlin and Livermore EBITs)
did not change the observed high hardness ratio. Another possibility
currently being investigated at Livermore is a change in ion temperature
during the switch from electron trapping mode to the
magnetic mode.

The difference in hardness ratios may also be caused by a large population in
high-$n$ metastable levels. Decay from those levels would not be observed in a
crossed-beam experiment if the metastable ions exit the
observation region before radiative decay can take place.
A search for such levels using atomic structure calculations
has not identified any possible candidate levels but those
calculations were for singly excited states, i.e., they assumed
single electron capture.  The question, therefore, is still
open whether there are metastable levels populated in multi-electron
capture events that would lead to delayed X-ray emission from high-$n$ levels.

\section{Summary and Future Work}
\label{sec:future}

Electron beam ion traps, first demonstrated twenty years ago,
have opened a new window on the study of charge exchange,
just as CX has been recognized as an important mechanism
in astrophysical X-ray emission.  The development of X-ray microcalorimeters,
which has proceeded roughly in parallel with that of EBITs,
likewise promises rapid advances in the study of CX
both in the lab and astrophysics.

As non-dispersive spectrometers with much better energy resolution than
Ge and Si(Li) solid state detectors, microcalorimeters planned for
use on rocket flights and future X-ray missions such as {\it Constellation-X}
will permit detailed study of diffuse CX emission sources.
At very low X-ray energies, specifically 
100 and 284 eV (corresponding to the \rosat\ 1/4-keV band), there
have been glimpses of CX emission seen in {\it Diffuse X-Ray Spectrometer}
(DXS; \cite{cit:sanders2001})
spectra, particularly the $\sim$67.4 \AA\ line, and by the
{\it X-Ray Quantum Calorimeter} (XQC; \cite{cit:mccammon2002}).
The DXS was a moderate-resolution scanning-mode crystal spectrometer that
flew on the Space Shuttle, and the XQC was a rocket-borne microcalorimeter.
Strong solar wind CX emission from L-shell ions such
as Li-like Mg, Si, and S is expected in the 1/4-keV band 
(and at higher energies), in accord with the observation that 
the Long Term Enhancements seen by \rosat\ were strongest in
that band \cite{cit:snowden1995}.  
Currently there are no laboratory or theoretical
data on such emission.

Another exciting opportunity for future research is the study of the
relative population of $S=1$ and $S=0$ states in He-like ions
formed by CX.  As noted in Section~\ref{sec:results-helike},
some work with Be-like ions (which, like He-like ions, have two valence
electrons) indicates population ratios significantly different
from the 3:1 expected from simple statistics, and
recent theoretical work (P.~Stancil, private comm.)
suggests that CX of H-like ions with He at low energies leads to initial
population ratios much larger than 3:1.
High-resolution EBIT spectra of He-like \ka\ emission, such as
shown in Figure~\ref{fig:HeFeCXvsDE}
will help resolve this question.

As discussed in Section~\ref{sec:results-targets},
many questions posed by multi-electron CX remain to be addressed.
Although EBITs cannot measure absolute CX cross sections, it should
be possible to derive relative cross sections for SEC and DEC in
some systems, as well as relative cross sections for CX
of bare and H-like ions of the same element, which will supplement the
limited data from low-energy crossed-beam experiments.

From an astrophysical perspective, H is the most important CX target.
Atomic hydrogen is difficult to produce and work with in the laboratory,
however, and
very few CX measurements with H have been published, none with X-ray
spectra at the several-eV/amu energies used in EBITs.
Experiments now underway with a new H source on the LLNL EBIT
will hopefully provide important tests of theoretical models, particularly
for higher-$q$ ions and low collision energies
that can not be studied with other ion sources.

\section*{Acknowledgements}

This work was supported by NASA's Space
Astrophysics and Analysis program under Grant NAG5-10443
and by NASA's Planetary Atmospheres Program under grant NNG06GB11G.
BW was also supported by NASA contract NAS8-39073 to
the \chandra\ X-Ray Center.
Work at the Lawrence Livermore National Laboratory
was performed under the auspices of the US Department of Energy under
contract No. W-7405-ENG-48.





\end{document}